\documentclass[useAMS,usenatbib]{mn2e}

\usepackage[latin1]{inputenc}
\usepackage{graphicx}
\usepackage{epstopdf}

\usepackage{subfigure}
\usepackage{rotate}
\usepackage{amsmath, float}
\usepackage{txfonts}
\usepackage{fleqn}
\usepackage{color}
\usepackage{comment}

\usepackage{mathrsfs}  
\usepackage{booktabs}
\usepackage{colortbl} 

\newcolumntype{G}{>{\columncolor[gray]{0.8}}l} 

\newcommand{\be}{\begin{equation}}
\newcommand{\ee}{\end{equation}}
\newcommand{\bdm}{\begin{displaymath}}
\newcommand{\edm}{\end{displaymath}}
\newcommand{\bea}{\begin{multline}}
\newcommand{\eea}{\end{multline}}
\newcommand{\ba}{\begin{align}}
\newcommand{\ea}{\end{align}}

\renewcommand{\sigma}{\varpi^2}

\def\simlt{\mathrel{\hbox{\rlap{\hbox{\lower4pt\hbox{$\sim$}}}\hbox{$<$}}}}
\def\simgt{\mathrel{\hbox{\rlap{\hbox{\lower4pt\hbox{$\sim$}}}\hbox{$>$}}}}

\title[NSs with twisted magnetosphere in GR]
{General relativistic neutron stars with twisted magnetosphere}
\author[A.~G. Pili, N. Bucciantini, L. Del Zanna]{
A.~G. Pili$^{1,2,3}$\thanks{E-mail: pili@arcetri.astro.it}, N. Bucciantini$^{2,3}$, L. Del Zanna$^{1,2,3}$ \\
$^{1}$Dipartimento di Fisica e Astronomia, Universit\`a degli Studi di Firenze, Via G. Sansone 1, 
I-50019 Sesto F.~no  (Firenze), Italy\\
$^{2}$INAF - Osservatorio Astrofisico di Arcetri, Largo E. Fermi 5, I-50125 Firenze, Italy\\
$^{3}$INFN - Sezione di Firenze, Via G. Sansone 1, I-50019 Sesto F.~no  (Firenze), Italy}

\begin{document}
 
\date{Accepted / Received}

\maketitle

\label{firstpage}

\begin{abstract}
Soft Gamma-Ray Repeaters and Anomalous X-Ray Pulsars are extreme manifestations
of the most magnetized neutron stars: magnetars. The phenomenology of their emission and
spectral properties strongly support the idea that the magnetospheres of these astrophysical
objects are tightly twisted in the vicinity of the star. 
Previous studies on equilibrium configurations have so far focused 
on either the internal or the external magnetic field configuration, without considering
a real coupling between the two fields. 
Here we investigate numerical equilibrium models
of magnetized neutron stars endowed with a confined twisted magnetosphere, 
solving the general relativistic Grad-Shafranov equation both in the interior 
and in the exterior of the compact object.
A comprehensive study of the parameters space is provided,
to investigate the effects of different current distributions on the overall magnetic field structure.

\end{abstract}


\section{Introduction}
Neutrons Stars (NSs) can manifest themselves as different 
classes  of astrophysical sources, each one of them characterized by a peculiar 
phenomenology.  Among these classes, Anomalous
X-Ray Pulsars (AXPs) and Soft Gamma-ray Repeaters (SGRs) are 
particularly remarkable because of their extraordinary energetic 
properties. Both exhibit a persistent X-ray emission 
with luminosities $L_X\sim  10^{33} - 10^{36}~\mbox{erg s}^{-1}$.
They are characterized by flaring activity with X-ray bursts whose
duration is  $\sim 0.1 -1$~s and with peak luminosities 
$\sim 10^{40}-10^{41} \mbox{ erg s}^{-1}$ which are,
in some cases, associated also with a pulsed radio transient emission.
 SGRs are sources of violent events, known as giant flares, 
during which an amount of energy $\sim 10^{44} - 10^{46}~\mbox{ erg}$ 
is released [for recent reviews see \citet{Mereghetti08a}, \citet{Rea_Esposito11a}
or \citet{Turolla_Esposito13a}].

There is a general consensus that SGRs and AXPs, on the ground of
their observational properties, are part of a same class of NSs
called \textit{magnetars} \citep{Duncan_Thompson92a,Thompson_Duncan93a}. 
These are young (with a typical age of $10^4$~yr), 
isolated NSs with rotational period in the 
range  $\sim 2 -12$~s, and with a typical dipole magnetic field,
inferred from spin-down, in the range $10^{14} - 10^{15}$~G \citep{Kouveliotou_Strohmayer+99a}.  
Since they are slow rotators, spin-down energy losses  cannot power their emission, 
which is instead believed to originate in the magnetic energy stored in the interior. 
The observed phenomenology would be thus sustained by 
the rearrangement and dissipation of their huge magnetic fields.

The simultaneous presence of high density, strong gravity, and strong magnetic
fields makes magnetars a unique environment.
Unfortunately, it is not yet known how such a strong magnetic field can form, 
and even less known are the requirements for its stability.
During the collapse of the progenitor star and the formation of the 
compact proto-NS, any fossil field, inherited by the
progenitor, can be either enhanced by the compression
of the core \citep{Spruit09a} or/and by its differential rotation. 
It is not clear, however, if this is enough to achieve the typical
magnetar field strengths, or if some form of dynamo is necessary
\citep{Thompson_Duncan93a,Bonanno_Rezzolla+03a,Rheinhardt_Geppert05a,Burrows_Dessart+07a}.  
In this case, proto-magnetars should be born as millisecond rotators, with
important consequences for the supernova explosion itself
\citep{Bucciantini_Quataert+09a,Metzger_Gioannios+11a,Bucciantini_Metzger+12a}.

In any case, there is no reason to expect that the magnetic field, at the very
beginning, is already in a stable configuration, while it is more likely expected that 
it will rapidly evolve into a stable one. This is because the Alfv\'enic crossing 
time is much smaller than the typical Kelvin-Helmholtz time-scale
\citep{Pons_Reddy+99a}. After $\sim 100$~s since the formation 
of the proto-NS, the neutrino driven wind ceases. This is the typical
time when an approximatively force-free
magnetosphere can be established outside the star.  It is also the time
when a crust begins to form, eventually freezing
the magnetic field on the NS surface. From now on the 
magnetic fields evolve on the much longer diffusive time-scale 
\citep{Braithwaite_Spruit06a,Vigano_Rea+13a,Gourgouliatos_Cumming+13a}.

The study of the properties of magnetic configurations inside NSs is
thus an important step towards a complete understanding of magnetars
and their properties. 
It is well known that either purely poloidal or purely toroidal magnetic 
configurations are  highly unstable 
\citep{Prendergast56a,Lander_Jones11a,Lander_Jones11b,Ciolfi_Rezzolla12a}.
However, given their simplicity, such kind of configurations have been extensively 
investigated in the context of equilibrium models of magnetized
NSs either in the Newtonian regime, from the earlier studies of 
\citet{Chandrasekar_Fermi53a} to
the more recent \citet{Yoshida_Yoshida+06a}, or in  full
General Relativity (GR) by \citet{Bocquet_Bonazzola+95a},
\citet{Kiuchi_Yoshida08a}, \citet{Frieben_Rezzolla12a} and \citet{Pili_Bucciantini+14a}, hereafter PBD14. 

Stability was investigated by  \citet{Braithwaite_Norlund06a} and 
\citet{Braithwaite09a} by means of numerical simulations in the 
Newtonian MHD regime.
The general outcome of the relaxation of the initial
magnetic field is a dynamically stable \textit{Twisted Torus} (TT) 
configuration, where the poloidal field is roughly 
axisymmetrically twined with a toroidal field, of comparable 
strength, into a ring-like region located just underneath the 
surface of the star. The exterior field they find is mainly a dipole with only 
smaller contributions from higher-order multipoles (note however that
they impose a potential field outside the NS surface, confining any
current to the interior). 

Given that TT configurations seem in principle to be stable,
particular efforts have been recently aimed at their investigation.
Models have been worked out in the Newtonian regime 
\citep{Lander_Jones09a,Glampedakis_Andersson+12a,Fujisawa_Yoshida+12a}, 
in GR with a perturbative approach \citep{Ciolfi_Ferrari+09a,Ciolfi_Ferrari+10a,Ciolfi_Rezzolla13a} 
and more recently also in the fully non-linear GR regime 
in PBD14.

The analysis of equilibrium
configurations has mainly focused on understanding the effects of 
the magnetic field on the structure of the star. Strong magnetic 
fields, indeed, could deform the star, and such deformations, in 
conjunction with fast rotation, could lead to emission of 
Gravitational Waves (GW) which, in principle, would be observed by the 
next generation of ground based detectors 
\citep{Mastrano_Melatos+11a,Lasky_Melatos13a}. This makes strongly magnetized
NS prime candidates for GW detection. Moreover,
there are some recent claims of a free precession in 4U~0142+61 
suggesting a prolate deformation of the NS, possibly
caused by a toroidal magnetic field of the order of $10^{16}$~G 
\citep{Makishima_Enoto14a}.

Until now, NS equilibrium models  have been developed
assuming that the star is surrounded by a
``vacuum''. However, since  the seminal work of 
\citet{Goldreich_Julian68a}, it is clear
that rotating NSs are actually surrounded by a magnetosphere filled by a charge-separated,
low-pressure plasma. Since the plasma pressure and its 
mass density are negligible with respect to
the electromagnetic energy density, the magnetosphere can be described
in the force-free approximation 
\citep{Contopoulos_Kazanas99a,Spitkovsky06a,Kalapotharakos_Contopoulos09a,
Contopoulos_Kalapotharakos+14a,Philippov_Tchekhovskoy+14}.

In the case of magnetars we expect that their magnetosphere is 
endowed with a high twisted magnetic field.
This is strongly suggested by the features of their persistent X-ray spectra, which
are well fitted by a blackbody-like component at $kT\sim 0.5$ keV, joined 
with a  power-law tail that becomes dominant above 10 keV \citep{Kuiper_Hermsen+2006a}. 
The latter  can be explained in terms of resonant cyclotron
scattering of the thermal photons by magnetospheric particles as proposed in 
\citet{Thompson_Lyutikov+02a} and \citet{Beloborodov_Thompson07a}. 
These authors pointed out that the dissipation of magnetic energy 
inside the star could induce a twist of the emergent magnetic field 
into a non-potential state which is sustained by electric currents 
that, threading the magnetosphere, might interact 
with the thermal photons emitted by the surface of the NSs.

The twisted magnetosphere scenario has been successfully validated
by calculation of synthetic spectra
\citep{Lyutikov_Gavriil06a,Fernandez_Thompson07a,Nobili_Turolla+08a,Taverna_Muleri+14a}. 
These works show that the morphology  of the external magnetic field, 
and the related charge distributions, highly affect 
the spectral shape and the pulse profile of the emitted radiation.
This indicates the importance of a correct modelling
and understanding of global magnetic configurations.
Typically, the standard reference model is the one discussed in 
\citet{Thompson_Lyutikov+02a} where, extending previous works on the
solar corona \citep{Low_Lou90a,Wolfson95a} to NSs, the magnetosphere 
is described in terms of a self-similar, globally twisted, dipolar magnetic field. 
This model has been refined to account for higher order multipoles by  
\citet{Pavan_Turolla+09a}, in response to  observational indications of 
a local, rather than global, twist in the magnetosphere 
\citep{Woods_Kouveliotou+07a,Perna_Gotthelf08a}. 
Recently this scenario has been strengthened also by the detection
of a proton cyclotron feature in the X-ray spectrum of the
``low-field'' magnetar SGR 0418+5729 which is compatible with
a strong, but localized, toroidal field of the order
of $10^{15}$ G \citep{Tiengo_Esposito+13a}.
More general equilibrium models have been obtained by 
\citet{Vigano_Pons+11a} using a magneto-frictional method, also first developed in 
the context of the solar corona studies \citep{Yang_Sturrock+86a,Roumeliotis_Sturrock+94a}, 
or by \citet{Parfrey_Beloborodov+13a} who, with  time-dependent numerical 
simulations, investigated the response of the magnetosphere to
different shearing profiles of the magnetic footpoints.

Given the complexity of the problem, until very recently
studies have focused either on the internal field structure (assuming
a prescription for the magnetosphere) or on the 
external magnetosphere (assuming an internal current distribution).
It is obvious that the two cannot be worked out
independently, and global models are the  necessary step forward.
A first attempt in this direction has been recently made in
\citet{Glampedakis_Lander+2013a}. Considering non-rotating 
stars in Newtonian regime, they show that a ``Grad-Shafranov approach'' 
to the problem can be used to obtain global equilibrium
configurations, with twists and currents extending from the interior
to the magnetosphere. 
A different approach was used in \citet{Ruiz_Paschalidis+14a}
where, for the first time,  detailed GR models of pulsar magnetosphere
were developed. In particular they search for steady state 
configurations by evolving in time the NS and by matching the  
interior field, treated with ideal MHD equation, to the exterior force-free solution.
 
In this work we extend previous results, and present for the first
time GR models of NSs endowed with a twisted magnetic field,
threading both the interior and the outer magnetosphere. 
Models are derived from the solution of the general-relativistic 
Grad-Shafranov equation, both in the interior and in the exterior
of the star. Our results are a generalization of the 
TT models presented in PBD14, that allow 
electric currents to flow outside the star. We have investigated
several models, varying either the strength of the currents producing the twist in 
the magnetosphere, or the extent of the magnetosphere itself. Extending the work of
\citet{Glampedakis_Lander+2013a}, where a couple of typical configurations
were presented, we develop a detailed study of the parameter space.  We investigate
how currents are distributed and how they affect the topology
of the resulting magnetic field. We show the modifications expected on the shape of 
the field at the surface, the magnetic dipole moment, and the energy stored 
in the toroidal component of the magnetic field.

The paper is organized as follows.
In Sect.~\ref{sec:math_model} we present the mathematical framework
adopted, and our choice for the currents defining the magnetosphere.  
In Sect.~\ref{sec:numerical} we describe the numerical set-up. 
In Sect.~\ref{sec:models} we present and discuss our models, and
finally we conclude in Sect.\ref{sec:conclusion}.
In the following we assume a signature $(-,+,+,+)$ for the spacetime
metric, employing Latin letters $i, j, k, \ldots$ (running from 1 to 3) for
3D spatial tensor components. We set $c=G =1$ and
all $\sqrt{4\upi}$ factors are absorbed in the definition of the electromagnetic fields.

\section{Formalism}

\label{sec:math_model}

All our magnetized NS models will be assumed here as non-rotating 
(as previously discussed, magnetars have a long rotation period) and axisymmetric. 
Ideal, General Relativistic Magnetohydrodynamics (GRMHD) is supposed to hold 
in the interior of the star, and it is also assumed to hold in the external magnetosphere,
where plasma inertia is certainly negligible (this actually corresponds to 
the so-called force-free regime).
Our formalism follows  the notation used in PBD14, to which the reader
is referred for a more complete discussion.

\subsection{The GRMHD Grad-Shafranov equation}

When the stress-energy tensor describing the matter distribution and the
magnetic field of a NS is axisymmetric, then the spacetime itself
must retain the same symmetry \citep{Carter70a,Carter73a}. 
As we showed in PBD14, NS models in full GR can be conveniently computed,
preserving a high accuracy, in the so-called \textit{conformally flat approximation} for
the spacetime metric \citep{Wilson_Mathews03a,Wilson_Mathews+96a}. 
This allows one to notably simplify Einstein equations
recasting them in a numerical stable form, and yet to derive results
that are fully consistent (with typical relative errors $\sim 10^{-4}$)
with more sophisticated approaches to GR 
\citep{Bucciantini_Del-Zanna11a,Cordero-Carrion_Cerda-Duran+09a}.
In the case of a static nonrotating star the line 
element of a conformally flat spacetime is written, using spherical 
like coordinates $(t,r,\theta,\phi)$, as
\be
\rmn{d} s^2 = - \alpha^2 \rmn{d} t^2 + \psi^4 (\rmn{d} r^2+ r^2 \rmn{d}\theta^2+r^2 \sin^2 \! \theta\, \rmn{d}\phi^2),
\label{eq:iso}
\ee
where $\alpha$ is the \textit{lapse function} and $\psi$ is the 
\textit{conformal factor}, dependent on the position.
We note that, since we consider a nonrotating star, the line element 
does not contain any mixed term $\rmn{d}x^i\rmn{d}t$, 
corresponding to a vanishing \textit{shift vector} $\beta^i=0$ in the 
$3+1$ formalism. Both the metric functions $\alpha$ and $\psi$ are
obtained solving Einstein equations that, in the specific case
of conformal flatness and a static NS, reduce to a set of 
two non-linear elliptic PDEs [see PBD14 or \citet{Bucciantini_Del-Zanna11a}].

In an axisymmetric and static spacetime the electromagnetic field can be described 
uniquely in terms of a magnetic potential, which coincides with the covariant $\phi$-component
of the vector potential $A_\phi$, and it is usually referred to as the \textit{magnetic flux
function}. In particular, the solenoidality condition, together with  axisimmetry,
allows one to express the poloidal component of the magnetic field as a gradient of the magnetic flux
function, whereas the toroidal counterpart is related to $A_\phi$ by means of a 
free scalar \emph{current function} $\mathcal{I}$ that depends on $A_\phi$ alone. 
Thus, under the assumption of a conformally flat metric, the components of the 
magnetic field are given by
\be
B^r = \frac{ \upartial_\theta A_\phi}{\psi^6 r^2\sin\theta}, \quad
B^\theta = -  \frac{ \upartial_r A_\phi}{\psi^6 r^2\sin\theta},\quad
B^\phi = \frac{\mathcal{I}(A_\phi)}{\alpha\psi^4 r^2\sin^2\theta}.
\label{eq:aphi}
\ee
From the static GRMHD system, in the presence of an external magnetic field
and assuming a barotropic Equation of State (EoS) for the fluid, the Euler equation can be written as
\be
\upartial_i \ln h + \upartial_i \ln\alpha  = \frac{\rmn{d}\mathcal{M}}{\rmn{d} A_\phi}\upartial_i A_\phi \, ,
\label{eq:euler}
\ee
where $\rho$ is the rest mass density, $h:=(e+p)/\rho$ is the specific enthalpy,
$e$ and $p$ are the energy density and the thermal pressure, respectively.
Here we have already related the Lorentz force component $L_i$ to the
gradient of the \textit{magnetization function} $\mathcal{M}(A_\phi)$ through
\be
\rho h \upartial_i \mathcal{M} = L_i = \epsilon_{ijk}J^j B^k \, ,
\label{eq:lorentzf}
\ee
in which $J^i= \alpha^{-1} \epsilon^{ijk} \upartial_j(\alpha B_k)$ are the 
conduction currents that can be expressed, in terms of $\mathcal{I}$ and
$\mathcal{M}$, with
\be
J^r = \alpha^{-1} B^r \frac{\rmn{d}\mathcal{I}}{\rmn{d} A_\phi}, \quad
J^\theta = \alpha^{-1} B^\theta \frac{\rmn{d}\mathcal{I}}{\rmn{d} A_\phi},\quad
J^\phi = \rho h \, \frac{\rmn{d}\mathcal{M}}{\rmn{d} A_\phi} + 
\frac{\mathcal{I}}{\sigma}\frac{\rmn{d}\mathcal{I}}{\rmn{d} A_\phi} ,
\label{eq:cur}
\ee
where we have defined $\varpi := \alpha \psi^2 r\sin\!\theta$.

Integrating equation~\eqref{eq:euler} one obtains the \textit{Bernoulli integral}
\be
\ln{\left( \frac{h}{h_c}\right)} + \ln{\left( \frac{\alpha}{\alpha_c}\right)} - \mathcal{M} = 0,
\label{eq:bernoulli}
\ee
which, once the functional form of $\mathcal{M}$ has been chosen
and both $\alpha$ and $A_{\phi}$ are available, relates the enthalpy at each point
to the conditions set at the centre of the star (labeled $c$), where we assume $\mathcal{M}_c=0$.
Finally, the magnetic flux function $A_\phi$ is related to the metric terms and the hydrodynamical
quantities through the GRMHD \emph{Grad-Shafranov} (GS) equation
\be
\tilde{\Delta}_3 \tilde{A}_\phi
+  \frac{\upartial A_\phi \upartial\ln (\alpha\psi^{-2})}{r \sin\theta}
+ \psi^8 r \sin\!\theta \left( \rho h \frac{\rmn{d} \mathcal{M}}{\rmn{d} A_\phi}
+ \frac{\mathcal{I}}{\sigma}\frac{\rmn{d}\mathcal{I}}{\rmn{d}A_\phi} \right) = 0.
\label{eq:gs}
\ee
This is obtained by working out the derivatives of the magnetic field in \eqref{eq:lorentzf} introducing,
for convenience, the new variable $\tilde{A}_\phi=A_\phi / (r\sin\theta)$
and the following differential operators
\be
\tilde{\Delta}_3  \! := \!  \Delta - \frac{1}{r^2\sin^2\!\theta}  \! = \! 
\upartial^2_r + \frac{2}{r}\upartial_r+\frac{1}{r^2}\upartial_\theta^2
+ \frac{1}{r^2\tan{\theta}}\upartial_\theta - \frac{1}{r^2\sin^2\!\theta},
\ee
\be
\upartial f\upartial g := \upartial_r f \upartial_r g+\frac{1}{r^2}\upartial_\theta f \upartial_\theta g .
\ee
The GS equation, which governs the GRMHD equilibrium inside the star, can be
extended also outside if we describe the external magnetosphere as a low density
plasma where the force-free regime is valid. Indeed, in the non-rotating case,
the force-free condition reduces to the vanishing of the Lorentz force $L_i=0$, and one
can again obtain a GS equation that corresponds to equation~\eqref{eq:gs} in the $\rho \rightarrow 0$
limit. We notice that taking the non-relativistic limit of equation~\eqref{eq:gs} with
$\rho \rightarrow 0$ leads to the non-rotating limit of the so-called
pulsar equation \citep{Glampedakis_Lander+2013a}.

Finally we recall that our choice of expressing
all the electromagnetic quantities as 
functions of $A_\phi$ is not appropriate in the case of a purely toroidal magnetic field, 
which instead requires a different description, see PBD14 or \citet{Gourgoulhon_Markakis+11a} for details.

\subsection{Choice of the free functions}
\label{sec:freefun}

The current and magnetization free functions entering the GS equation can be easily modified,
with respect to the choice made in PBD14, in order to allow the currents
to flow also outside the star. 
We can actually retain the same form for the magnetization function $\mathcal{M}$
\be
\mathcal{M}(A_\phi)=k_{\rm pol} A_\phi ,
\label{eq:mbern}
\ee
where  $k_{\rm pol}$ is the \emph{poloidal magnetization constant}, while we adopt
here a different functional form for $\mathcal{I}$, namely
\be
\mathcal{I}(A_\phi)=\frac{a}{\zeta+1}\Theta [A_\phi-A_\phi^{\rm ext}] 
\frac{(A_\phi-A_\phi^{\rm ext})^{\zeta+1}}{{(A_\phi^{\rm max})}^{\zeta+1/2}},
\label{eq:fbern}
\ee
where $\Theta [.]$ is the Heaviside function, $A_\phi^{\rm max}$ is
the maximum value that the $\phi$ component of the vector potential reaches over the entire
domain, while $A_\phi^{\rm ext}$ is the maximum value it reaches
at a distance $r=\lambda r_{\rm e}$ from the star (being $r_{\rm e}$  the equatorial
radius). We further define $a$ as the \emph{toroidal magnetization constant}, whereas
$\zeta$ is the \emph{toroidal magnetization index}. Note that these choices are
analogous to the ones made in \citet{Glampedakis_Lander+2013a}.
The new parameter $\lambda$, that enters in the definition of  $A_\phi^{\rm ext}$,
allows us to control the size of the twisted magnetosphere outside the star.
The results of PBD14 are recovered assuming $\lambda=1$. 
On the other hand, for $\lambda >1$ the toroidal magnetic 
field is not confined within the star but extends smoothly outside the 
stellar surface, just like the poloidal component.

From the relations \eqref{eq:cur}, given our choice for the
free functions $\mathcal{M}$ and $\mathcal{I}$, 
the components of the conduction currents become
\be
J^r = \alpha^{-1} B^r \, a\Theta[A_\phi-A_\phi^{\rm ext}]
\frac{(A_\phi-A_\phi^{\rm ext})^{\zeta}}{{(A_\phi^{\rm max})}^{\zeta+1/2}}, 
\ee
\be
J^\theta  = \alpha^{-1} B^\theta \, a\Theta[A_\phi-A_\phi^{\rm max}]
\frac{(A_\phi-A_\phi^{\rm ext})^{\zeta}}{{(A_\phi^{\rm max})}^{\zeta+1/2}},
\ee
\be
J^\phi = \rho h \, k_{\rm pol} + 
\frac{a^2}{(\zeta+1)\sigma} \Theta[A_\phi-A_\phi^{\rm ext}]
\frac{(A_\phi-A_\phi^{\rm ext})^{2\zeta+1}}{{(A_\phi^{\rm max})}^{2\zeta+1}}.
\label{jphic}
\ee
Note that, thanks to the renormalization by $A_\phi^{\rm max}$,
the $\phi$ component of the conduction current is independent from the
absolute value of the magnetic flux function (it does not depend on the
field strength), while it is  directly controlled only by the 
magnetization parameter $k_{\rm pol}$ and $a$.

In the low-magnetization limit, when the magnetic energy $\mathcal{H}$
is smaller than the gravitational mass $M$ ($\mathcal{H}<<M$), the metric functions $\alpha$ 
and $\psi$ depend weakly on the magnetic field strength, and one can
safely assume for them the same values of the unmagnetized case. We
have verified that the low-magnetization limit applies as long as the
magnetic field at the centre is weaker than  $10^{16}$~G
(corresponding to a magnetic field at the surface smaller than a few
$10^{15}$~G). For smaller values, non-linear variations are absent
(changes in the results are well within the overall accuracy of the
scheme), while they become evident at higher values.  The reader is
referred to Appendix~\ref{sec:app1} for a more detailed  discussion 
on the strong field regime.

Interestingly, in this limit, it is possible to recast the current function
$\mathcal{I}$ in a self-similar way, such that the resulting magnetic
field configuration remains unchanged, modulo its strength.  
If $J^\phi$ is rescaled with a numerical factor $\eta$
sending $k_{\rm pol}\mapsto \eta k_{\rm pol}$ and $a \mapsto \sqrt{\eta} a$, the solution
of the GS equation itself is rescaled by the same numerical
factor. The self similar parameter can be thus defined in terms of the strength
of the magnetic field as
\be
\hat{a}=a\left(\frac{B_{\rm pole}}{10^{14} {\rm G}}\right)^{-\frac{1}{2}},
\ee
where $B_{\rm pole}$ is the magnetic field at the pole, that we have
decided to always normalize against $10^{14}$~G. Then the quantity $\hat{a}$
parametrizes the magnetic configurations, independently from the
strength of the magnetic field. Notice that in PBD14, since  the choice for the
current function $\mathcal{I}$ in the TT case had a slightly different
normalization , the role of self-similar parameter was assumed by $a$.
Here the new normalization improves the convergence of our scheme preventing
the non-linear term from diverging at the highest value for $a$ and $\lambda$. Moreover it
allows us to obtain also configurations with a complex magnetospheric 
field geometry (see section \ref{sec:models}). 
 In this work we will focus exclusively on the 
low-magnetization limit, considering for simplicity cases with $\zeta=0$ or $\zeta=1$.

\section{Numerical setup}
\label{sec:numerical}

The numerical scheme used to compute our models is fully described 
in \citet{Bucciantini_Del-Zanna13a} and in PBD14, to which the reader
is referred for a more complete discussion. Here we quickly summarize
for convenience the main 
features and the few modifications introduced in the present work.

The basic idea of the algorithm is to use an expansion in spherical
harmonics to solve the non-linear Poisson-like equations, and reduce them
to a set of ordinary 2nd-order partial differential 
equations for each coefficient, that can be solved using a direct tridiagonal matrix inversion.
In the weak field limit the metric and matter distribution are assumed
to be the same as in the unmagnetized case, so that the we do not need to
solve the CFC Einstein equations and the problem is
reduced to find the solution of the GS equation~\eqref{eq:gs} alone,
a non-linear vector Poisson equation for $A_\phi$. 
Its solution is searched expanding $\tilde{A}_{\phi}$  by means of vector spherical
harmonics \citep{Barrera_Estevez85a}, that is
\be
\tilde{A}_\phi
(r,\theta):=\sum_{l=0}^{\infty}[C_l(r)Y^\prime_l(\theta)].
\label{eq:harmonics}
\ee 

As already done in PBD14 the solution to the GS equation is searched over the entire
numerical domain, which includes both the interior of the star and  the surrounding magnetosphere 
where the density, for numerical reasons, is set to a very small value
(in principle it could be set to 0). With this approach, there is no need to match the exterior
solution with the interior one (as it is usually done when the solution
of the GS equation is separately computed over disconnected domains) 
and the smoothness of the solutions at the stellar 
surface is here automatically guaranteed avoiding the onset of spurious surface currents.
The harmonic decomposition ensures the correct behaviour of the solution
on the symmetry axis. At the centre of the star the radial coefficients $C_l(r)$ go to
0 with parity $(-1)^l$, while a correct asymptotic trend at larger radii
(the outer boundary condition) is achieved by imposing $C_l(r)\propto r^{-(l+1)}$.

Being interested only in the study of the properties and geometry of
the magnetic field, in all our models we assume the NS to be
described by 
a simple polytropic EoS $p=K_a \rho^{\gamma_a}$ with an adiabatic index ${\gamma_a}=2$ and 
a polytropic constant, expressed in geometrized units, $K_a=110$.
This is done in analogy
with PBD14 and according to common choices in literature \citep{Kiuchi_Yoshida08a,Lander_Jones09a}.
Our fiducial model has a central density $\rho_c=8.576 \times 10 ^{14} \, \mbox{g} \, \mbox{cm}^{-3}$  a
baryonic mass $M_0=1.680 M_{\sun}$, a gravitational mass $M=1.551 M_{\sun}$, 
and a circumferential radius $R_{\rm circ}=14.19 \, \mbox{km}$. For
convenenience the magnetic field at the pole is rescaled to $B_{\rm
  pole}=10^{14}$~G (recall that results in the weak field limit are actually
independent from the field strength).

To explore the parameter space we computed several equilibrium models 
for different values of the parameter $\lambda$ (from $\lambda=1$ to $\lambda=8$)
and for different values of the parameter $\hat{a}$.

The numerical solutions we present here are computed using 60
harmonics. Models with  $\lambda < 4$ are computed over a uniform grid in spherical coordinates
covering the range $r=[0,40]$ and $\theta=[0,\upi]$ with 600 grid points in the radial direction and 400 
points in the angular one.
Models with $\lambda \geq 4$ have a twisted magnetosphere extending to
larger radii. In order to retain the same accuracy in the inner
region, and to reduce the computational time, we adopt a geometrically
stretched grid in the range $r=[40,150]$ defined on 200 grid points. 
The grid spacing $\Delta r$  is chosen such that
\be
\Delta  r_i=(1+5.962\times 10^{-3})\Delta r_{i-1}.
\ee
This permits to capture the entire twisted magnetosphere and to resolve the star always with the same
accuracy,  without resorting to huge numerical grids. 
The convergence tolerance for the iterative solution of
Grad-Shafranov equation as been fixed to $\sim 10^{-8}$,
however we have verified that the overall accuracy of our solution
are $\lesssim 10^{-3}$ because of the discretizations errors.

\section{Results}
\label{sec:models}

In this section we present the results of the GRMHD calculations.  
Since we focus on the low-magnetization limit, as discussed previously,
global physical quantities like the gravitational mass,
the baryonic mass, and the circumferential radius do not change for the various
sequences but remain equal to those of the fiducial model. Our
discussion will concentrate only on the magnetic properties of the
equilibrium configurations.  
All models are thus parametrized just in terms of $\hat{a}$, 
defining the magnetic field geometry, and $\lambda$, defining the 
extent of the magnetosphere.
For convenience magnetic field strengths are expressed in term
of their value at the pole $B_{\rm pole}$, that we arbitrarily assume to be $B_{\rm pole}=10^{14}$~G.

 In the following subsections we will consider only configurations with $\zeta=0$ and $\zeta=1$.
However a detailed investigation about the effects of more different and general current distribution for 
both $\mathcal{I}$ and $\mathcal{M}$ can be found in Bucciantini et al. 2014 (submitted).

\subsection{Models with $\zeta=0$}

\begin{figure*}
	\centering
	\includegraphics[width=.4\textwidth]{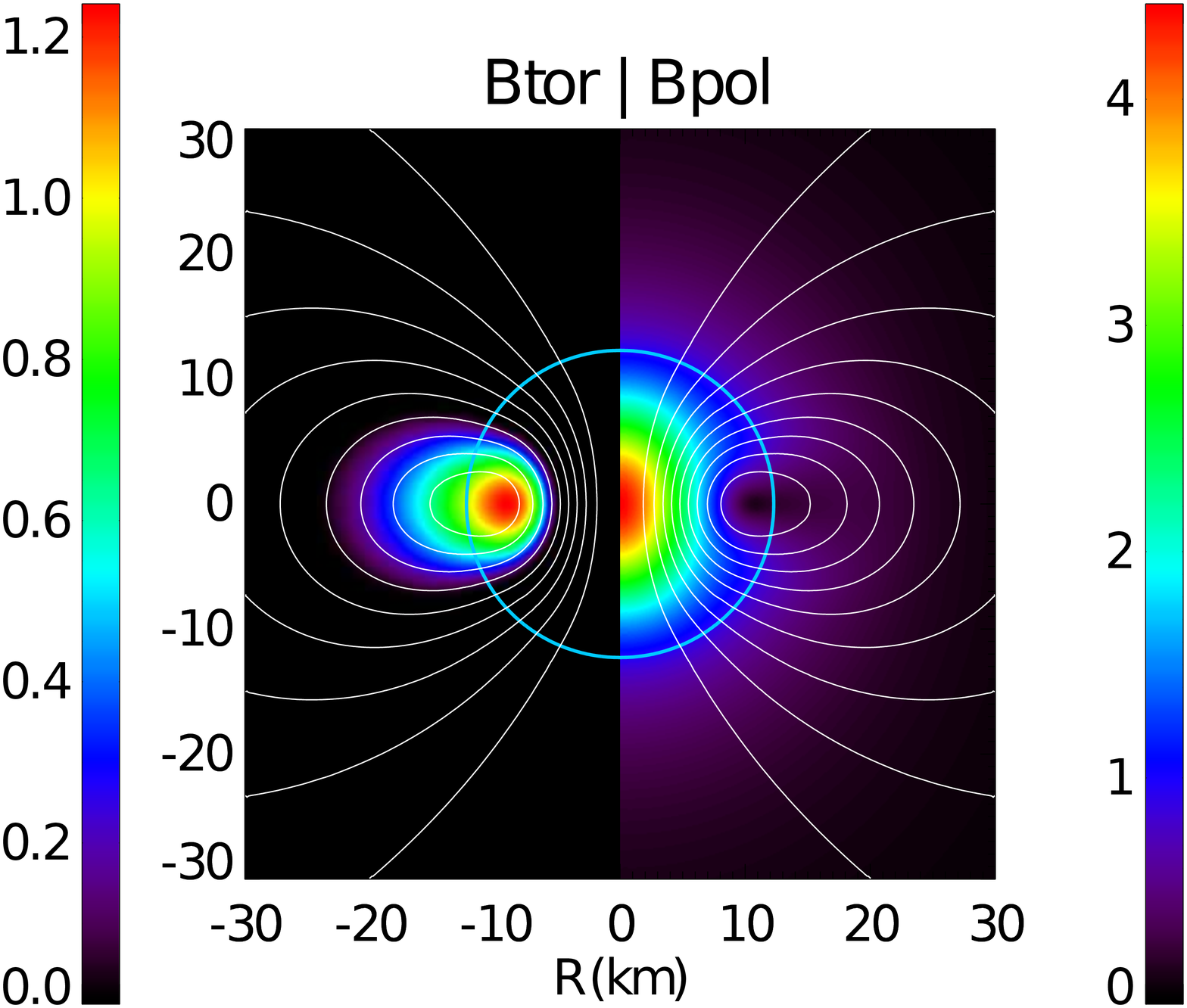} 
	\hspace{.05\textwidth}
	\includegraphics[width=.4\textwidth]{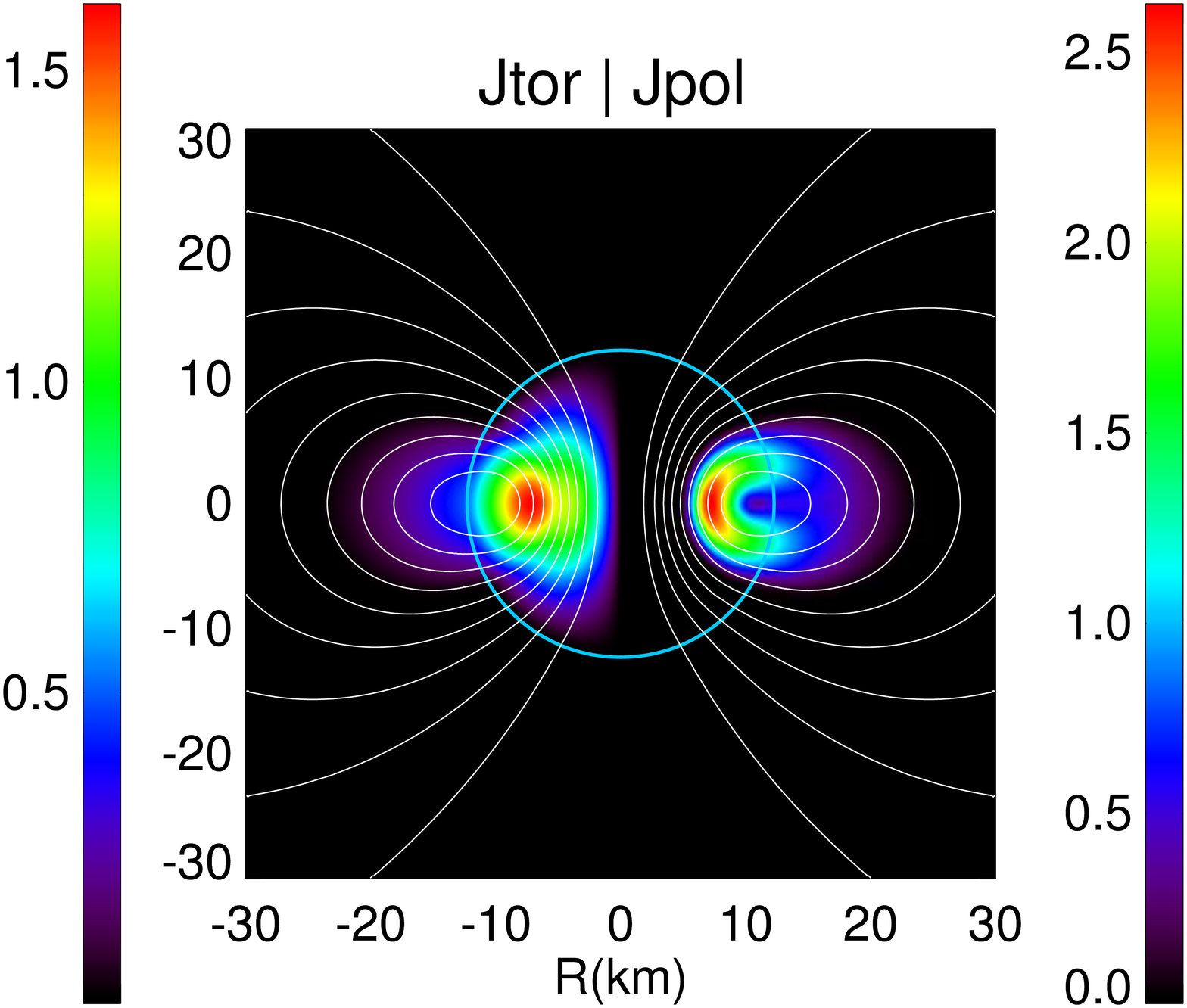} 
	\caption{ Left: strength of the toroidal (left half)  and poloidal (right
                  half) magnetic field in units of $B_{\rm pole}$. Contours represent magnetic
	          field surfaces. 
	          Right: same as the left panel for the toroidal  (left half)
                  and poloidal (right half) current density . The toroidal current density 
                  is expressed in units of
	          $10^{18} \, {\rm G} \, {\rm s}^{-1} $, the poloidal one in units of 
	          $10^{13} \, {\rm G} \, {\rm s}^{-1} $.
	          In both panels the blue curves represents the surface of the star. 
	          This configuration has $\lambda = 2$ and
	          $\hat{a}=2.5$, corresponding the highest value for the magnetic energy
	          ratio $\mathcal{H}_{\rm tor}/\mathcal{H}=11.29 \times 10^{-2}$.
              }
	\label{fig:maxratio}
\end{figure*}

In figure~\ref{fig:maxratio} we show a typical example of an equilibrium
model with a twisted magnetosphere. This specific configuration
corresponds to $\lambda=2$ and $\hat{a}=2.5$ . 
The poloidal magnetic field extends through the whole domain and 
reaches its maximum strength
$B_{\rm pol}^{\rm max}=4.422 \, B_{\rm pole} $
at the centre of the star. The toroidal component 
of the magnetic field is, by construction, confined inside a closed
region that extends in the radial direction from the interior of the star up to
twice the stellar radius, and in latitude it is contained within a wedge about 
$\pm \upi/6$ around the equator.  The maximum value of the
toroidal magnetic field $B_{\rm pol}^{\rm max} = 1.256 \, B_{\rm pole}$ is reached inside the star
in correspondence to the neutral line where the poloidal magnetic
field vanishes. The right panel of  figure~\ref{fig:maxratio} shows that the poloidal current 
density peaks inside the star, and extends smoothly outside
the stellar surface along the magnetic field surfaces. The toroidal current, on the other hand,
results from the sum of the linear 
current term in $\mathcal{M}$,  $J^{\phi}=\rho h k_{\rm pol}$, fully confined within the star, 
and of the non-linear term in $\mathcal{I}$,  that extends outside the star over the same
region where the poloidal currents are confined. 

The magnetospheric equilibria of the type shown in figure~\ref{fig:maxratio}, and discussed above,
are qualitatively similar to previous results \citep{Parfrey_Beloborodov+13a,Vigano_Pons+11a,Mikic_Linker94a}. 
However, in those cases the equilibria were
obtained  by the relaxation of an initially sheared dipolar
configuration, while here we directly solve the GS equation.
Such configurations, for a moderate shear of the magnetic footpoints, are
expected to be stable.  On the other hand, our approach based on the GS equation
allows us to derive equilibrium models but, of course, it does not provide any hint about their stability.
A more direct comparison can be made with \citet{Glampedakis_Lander+2013a},
in spite of a different value $\zeta=0.5$ employed, our solutions
qualitatively agree with the one presented in more detail in the cited work.
 
\begin{figure*}
	\centering
	\includegraphics[width=1.\textwidth]{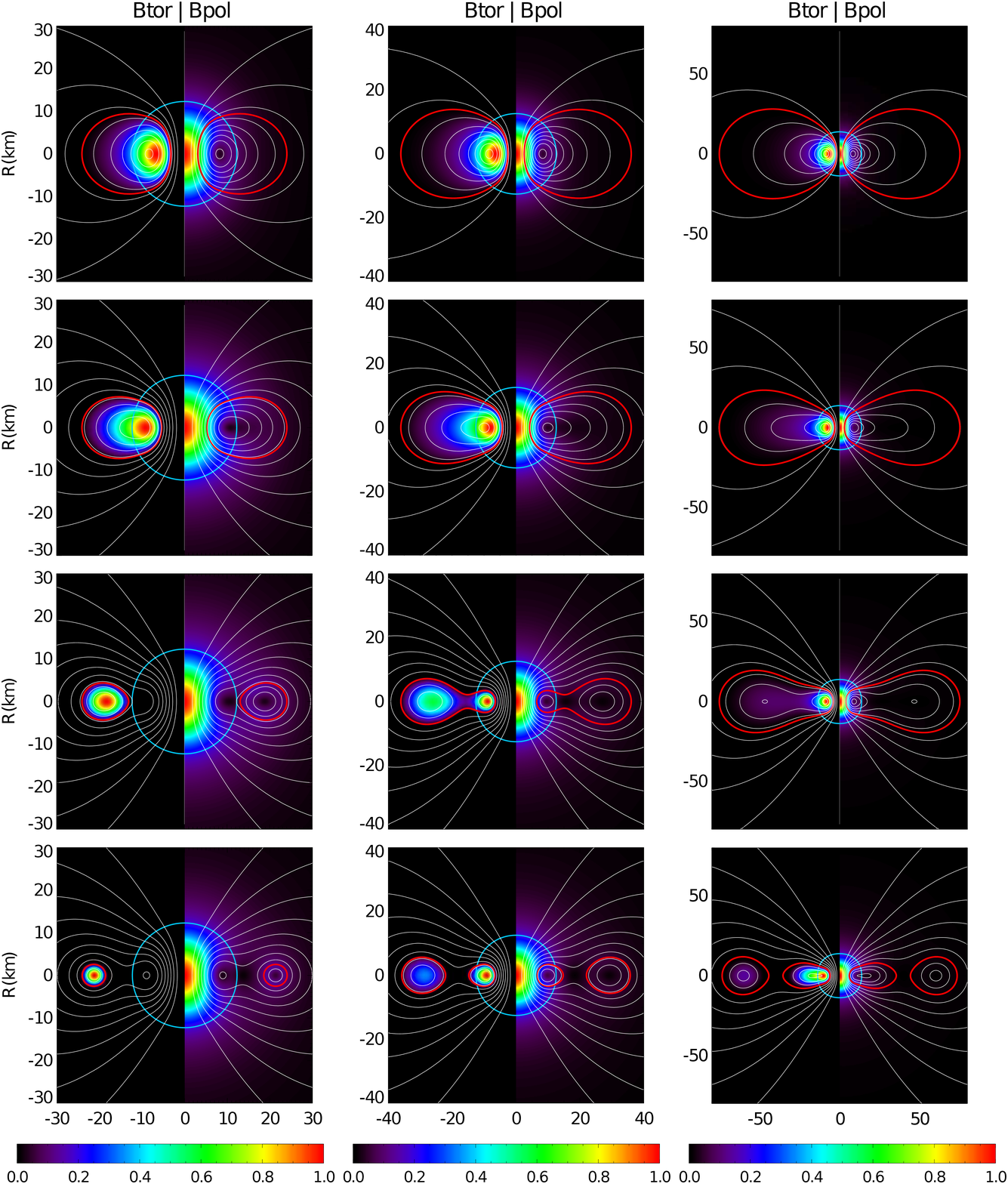}
	\caption{TT magnetosphere configurations:  
	strength of the toroidal (left half of each panel)  and poloidal (right
    half of each panel)  magnetic field.  The left column shows models
    with $\lambda=2$, the central column those with
    $\lambda=3$, and the right column those with $\lambda=6$.
	Contours represents magnetic field surfaces.
	From top to bottom each row corresponds to 
    increasing values of $\hat{a}$,given in Table~\ref{tab:collection}.
	For each panel the colour code is normalized to
	the maximum value of the magnetic field components
	that are listed in table~\ref{tab:collection}.
	The blue line represents the surface of the star.
	The red line locates the boundary of the region
	where the toroidal component of the magnetic
    field is present.
              }
	\label{fig:manymod}
\end{figure*}
\begin{table}
\centering
\caption{ \label{tab:collection}
Values of the maximum strength of the toroidal and poloidal components of the magnetic field
for the configuration shown in figure~\ref{fig:manymod}. 
Magnetic fields are expressed in unity of $B_{\rm pole}$.}
\begin{tabular}{l*{3}{c}}
\toprule
\toprule
$\lambda$ &  $\hat{a}$  & $B_{\rm tor}^{\rm max}$ &  $B_{\rm pol}^{\rm max}$ \\
	      & [$10^{-3}$] &            &             \\
\midrule
2		&	0.4   &   0.359  &  5.629 \\
		&	2.5	  &   1.256  &	4.422 \\
		&	5.6	  &	  0.938	 &  4.576 \\
		&   10    &   1.087  &   4.373 \\[.25cm]
		
3		&	0.4   &   0.435  &  5.581 \\
		&	1.5	  &   1.004  &  4.577 \\
		&	3.3	  &	  0.676	 &  4.767 \\
		&   4.2   &   1.170  &  4.618 \\[.25cm]
		
6		&	0.4   &   0.491  &  5.352 \\
		&	0.9	  &   0.782  &  4.874 \\
		&	1.3	  &	  0.822	 &  4.842 \\
		&   2.7   &   0.753  &  4.277 \\[.25cm]		
\bottomrule 
\end{tabular}
\end{table}

\begin{figure*}
	\centering
	\includegraphics[width=.33\textwidth]{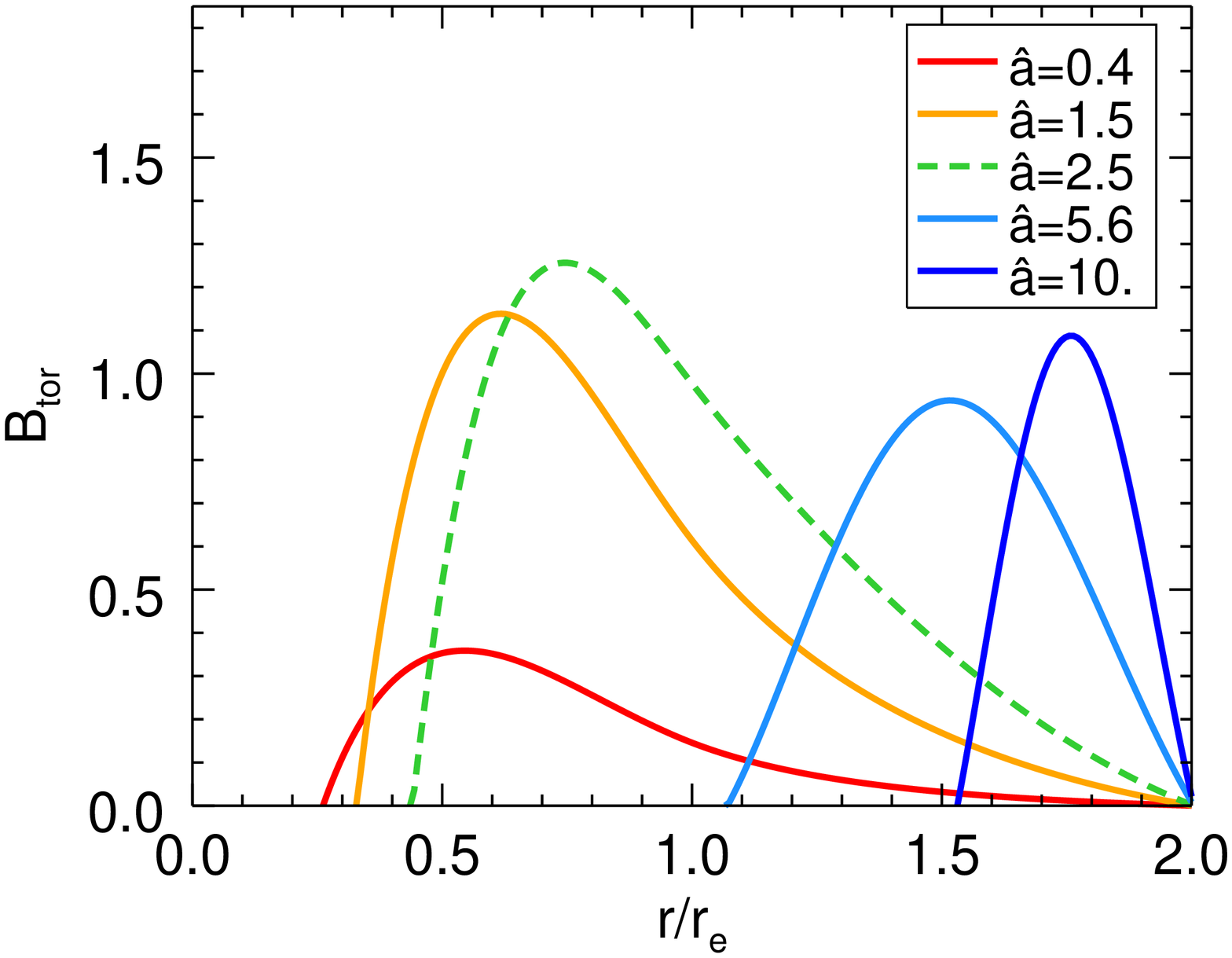}
	\includegraphics[width=.33\textwidth]{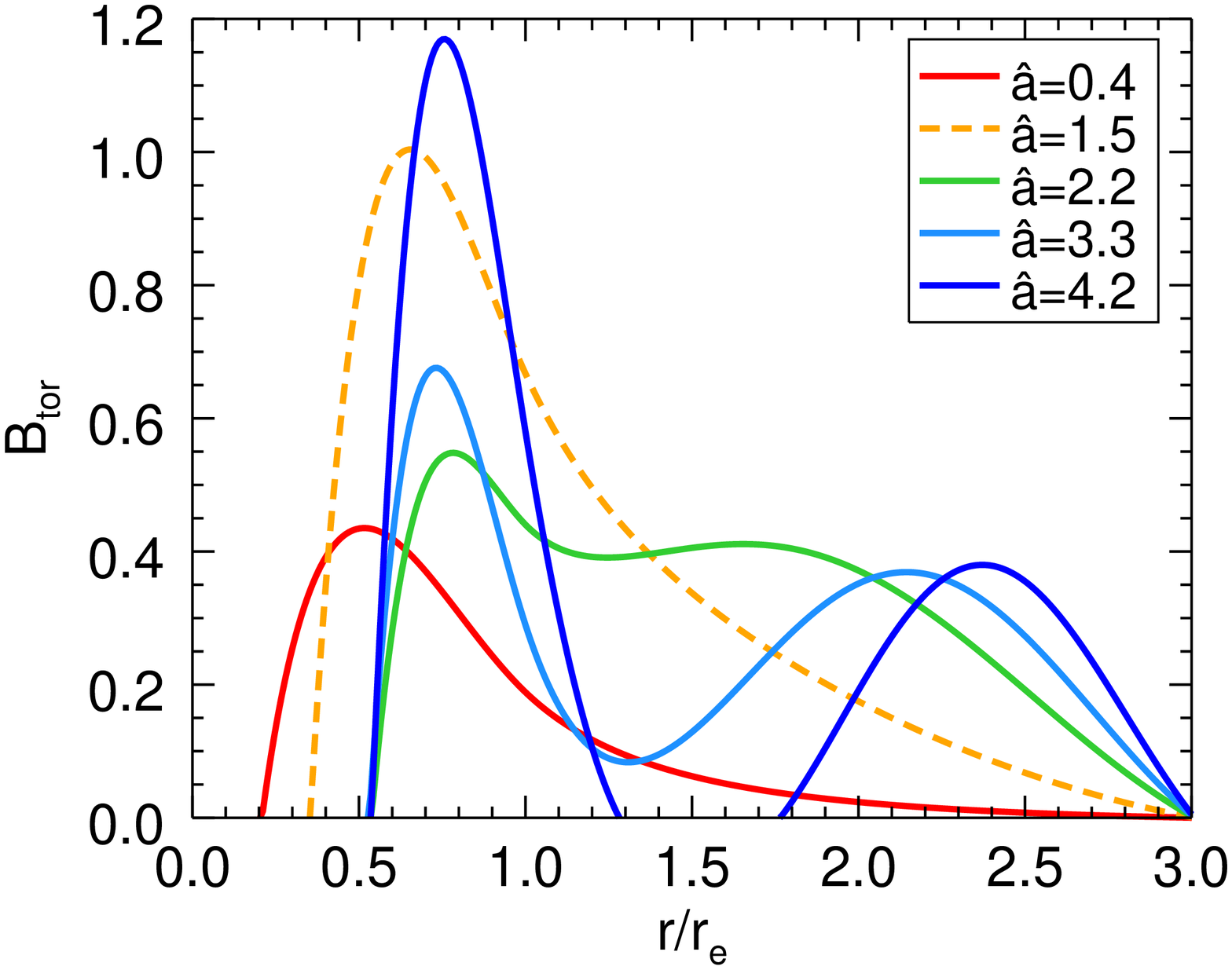}
	\includegraphics[width=.33\textwidth]{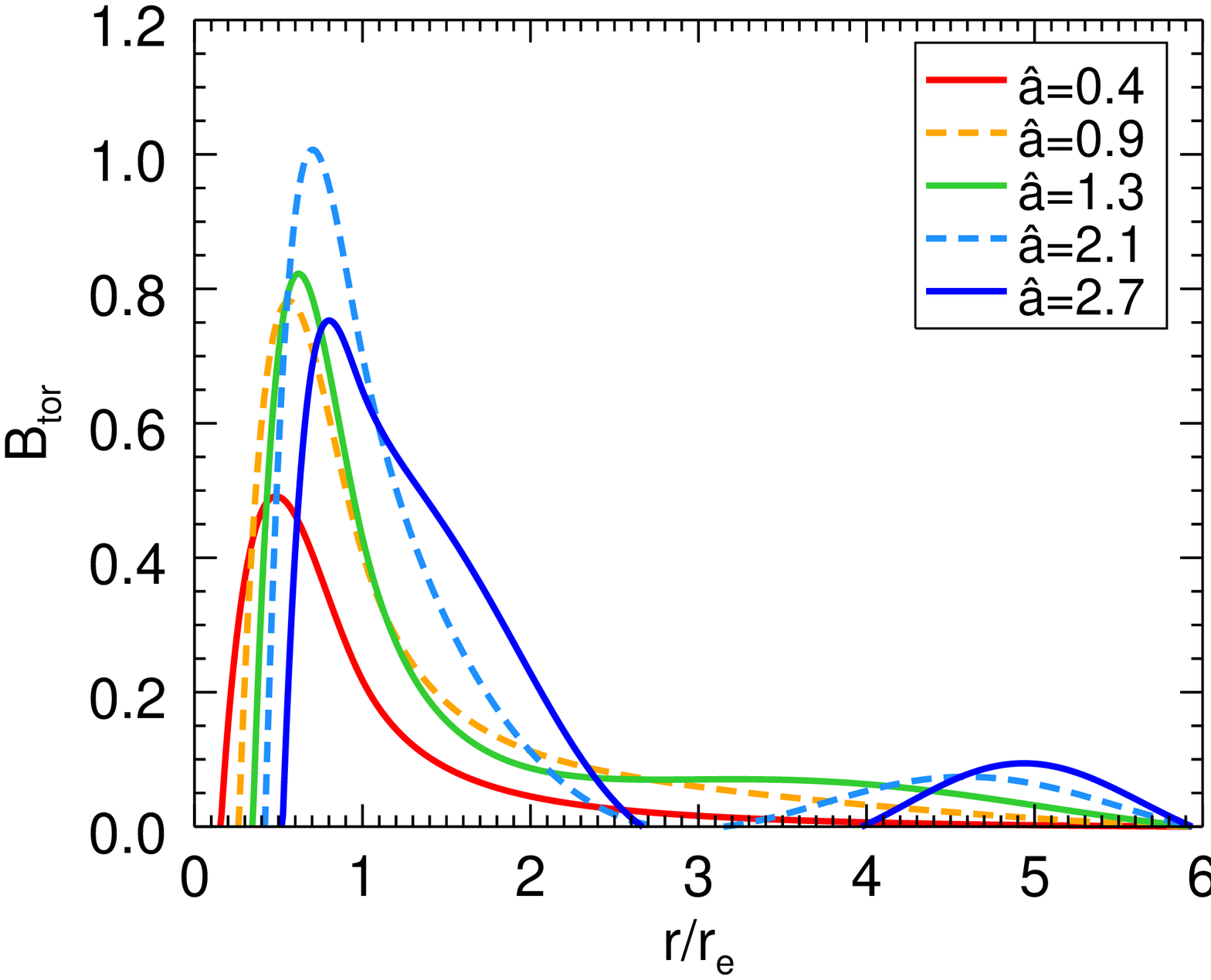}
	\caption{ Profiles of the toroidal magnetic field strength (in units of $B_{\rm pole}$) for selected 	models
	          ($\hat{a}$ is expressed in units of $10^{-3}$)
                  along the equilibrium sequences 
	          with $\lambda=2$ (left), $\lambda=3$ (centre) and  $\lambda=6$ (right).
	          The dashed lines represent models where the value of $\mathcal{H}_{\rm tor}/\mathcal{H}$
	          reaches a maximum.
	                        }
	\label{fig:bprofiles}
\end{figure*}

In figure~\ref{fig:manymod} we present sequences of models computed
for various values of the parameters $\lambda$ and $\hat{a}$. The main
characteristic of those configurations are stated in Table~\ref{tab:collection}.
The results presented illustrate the key features and trends of the
equilibrium configurations that we were able to obtain.
For small values of the parameter  $\hat{a}$ (first row of figure~\ref{fig:manymod}), all configurations
share the same overall topology and a similar magnetic field distribution.
The toroidal magnetic field fills a region that smoothly extends from
the interior of the star to the maximum allowed radius $\lambda r_e$. 
As it can be seen also from  figure~\ref{fig:bprofiles}, the toroidal
field component reaches, in this case, a maximum inside the star and
then decreases monotonically in the magnetosphere.
For small values of  $\hat{a}$, the non-linear current terms associated
to $\mathcal{I}$ outside the star are still too weak to significantly
alter the magnetic field structure below the surface. 

As  the contribution of the non-linear
currents become more important, with increasing $\hat{a}$, the toroidal field 
increases and its peak moves toward the stellar surface together with
the poloidal neutral line. This is a typical behaviour of TT
configurations, already observed in the case where the toroidal field
is fully confined within the star (PBD14; \citealt{Lander_Jones09a,Ciolfi_Ferrari+09a}). As the
toroidal magnetic field increases, the ratio of the magnetic
energy associated to the toroidal field $\mathcal{H}_{\rm tor}$ with
respect to the total magnetic energy
$\mathcal{H}$ increases too, until it reaches a maximum. 
These maximal configurations are shown in the second row of
figure~\ref{fig:manymod}. The structure and topology of the magnetic
field is analogous to the small $\hat{a}$ cases (see also figure ~\ref{fig:bprofiles}),
however the presence of stronger magnetospheric currents now affects the field geometry
outside the star. While the outer magnetic field in the small $\hat{a}$
regime still resembles closely a dipole, this is no longer true for the maximal energy
configurations, where magnetic surfaces appear to be stretched, especially
for high values of $\lambda$. Note, moreover, that for higher
$\lambda$, the configurations of maximum energy ratio are reached for 
smaller values of $\hat{a}$. This because the energy is an integrated
quantity that depends not just on the strength of the currents but also
on the volume they fill.

However, after the maximum value of $\mathcal{H}_{\rm tor}/\mathcal{H}$
has been reached, solutions react differently to a further increase
of $\hat{a}$, depending on the value of the parameter $\lambda$.
This can be seen in the third
row in figure~\ref{fig:manymod} and in figure~\ref{fig:bprofiles}. In the
case of $\lambda \le 2$ the toroidal magnetic field 
migrates completely outside the star and the final outcome strictly
resembles that of the TT case with $\lambda=1$ discussed in PBD14: 
the toroidal magnetic field strength grows but its support progressively shrinks 
toward the maximum allowed radius. Here the toroidal magnetic field shows a
single maximum. On the other hand, for $\lambda \ge 3$,
as the neutral line approaches the stellar surface, a second peak in
the strength of the toroidal magnetic field develops. This second peak moves with
increasing $\hat{a}$ at larger radii in the magnetosphere,
while the first peak remains inside the star, approximatively
at the same position, independently of $\hat{a}$. The formation of a
second peak indicates a topological change in the structure of
the magnetic field, where an X-point arises, usually in the vicinity
of the stellar surface, and where there are magnetic
regions (surfaces) in the magnetosphere disconnected from the star. A
further increase in the value of  $\hat{a}$ leads to solutions that
show two completely disconnected magnetic regions, one inside the star,
and the other outside (see the fourth row of figure~\ref{fig:manymod}). 
Note also that the maximum value of the strength of the toroidal
magnetic field $B_{\rm tor}^{\rm max}$ does not grow monotonically with $\hat{a}$.

These types of equilibria, with disconnected magnetic regions, are
likely to be highly unstable. Indeed,
those kind of equilibria resemble the solutions find in time-dependent numerical
simulation by \citet{Mikic_Linker94a}  in the context 
of magnetic field arcades in the solar corona, and \textit{plasmoid} formation. 
Indeed, our disconnected regions in the NS magnetosphere could
be seen as the equivalent of the plasmoids in the solar case.

While a full 3D study of the stability and/or evolution, of the various
  topological configurations, is beyond the scope of this paper, it is
  possible to roughly evaluate the magnetospheric conditions, in
  relation to known stability criteria. We need also to recall here
  that the physical regime, to which our models apply, is typical of
  the late phases of the proto-NS evolution, when a crust begins to form.

In all the obtained configurations the energy of the external toroidal magnetic field
is, at most,  25\% of the total magnetic energy in the magnetosphere which is, thus, 
dominated by the poloidal field. It is important, at this point, to distinguish between those configurations, where all
field lines thread the crust, and those with disconnected region. In the first case,
if the poloidal component can be stabilized by 
the crust (which can be the case for weak fields $\simlt 10^{14}$G),
then it is unlikely that the toroidal one, being subdominant, could
drive major changes in the magnetospheric structure. It is possible to
compare our results to those by  \citet{Parfrey_Beloborodov+13a},
where a study of the magnetospheric stability was done using a time
dependent shearing algorithm. We find in our models (those with no
disconnected regions) that the twist
amplitude, defined as the azimuthal angular displacement
of the magnetospheric footpoints,   does not exceed $2$ radians. This value is below the critical 
value of $3.65$ radians estimated in \citet{Parfrey_Beloborodov+13a}
as a stability limit for the magnetosphere.

In the other cases, when a disconnected toroidal current loop develops in the magnetosphere around
the neutral line, since some of the magnetic field lines do not cross the crust, the twist amplitude is 
not an indicative parameter for the stability, and one cannot invoke
for this disconnected region a stabilizing effect of the crust.
However the stability of this flux rope, can be determined from the Kruskal-Shafranov condition
for the development of kink instability \citep{Shafranov56,Kruskal_Tuck58}. The
value of the safety factor is $\sim 1$
 for detached flux ropes contained 
inside $1.3 r_{\rm e}$ (suggesting a possible marginal stability) but
rapidly drops to lower values as this disconnected region extends
further out from the star.
Therefore small disconnected magnetospheric regions just above the surface of the star appears more stable than inflated ones at larger distance.

The reason why the GS equation, for large $\hat{a}$,
admits solutions with multiple peaks can be easily understood. The
solution of the GS equation can be seen as an eigenvalue problem 
for  a second order non-linear PDE. For small values of
$\hat{a}$ the source terms in the currents are dominated by $\mathcal{M}$: they are
fully confined within the star and have a single peak. Thus the
solution reflects the properties of the source, and only single-peak
eigenmodes are selected. However, for higher
values of $\hat{a}$, non-linear terms dominates, and other possible
eigenmodes can be selected. Eigenmodes that for a second order
non-linear PDE, in principle will admit multiple radial nodes (this is
the reason why two peaks develop). Indeed, as can be seen from 
figure~\ref{fig:manymod} and \ref{fig:bprofiles}, there is some hint that the more 
extreme case at $\lambda=6$  might develop into a third peak. Unfortunately, we
could not investigate higher values of $\hat{a}$ because the convergence
of the GS solver becomes highly oscillatory, and ultimately fails. 

\begin{figure*}
	\centering
    \includegraphics[width=.33\textwidth]{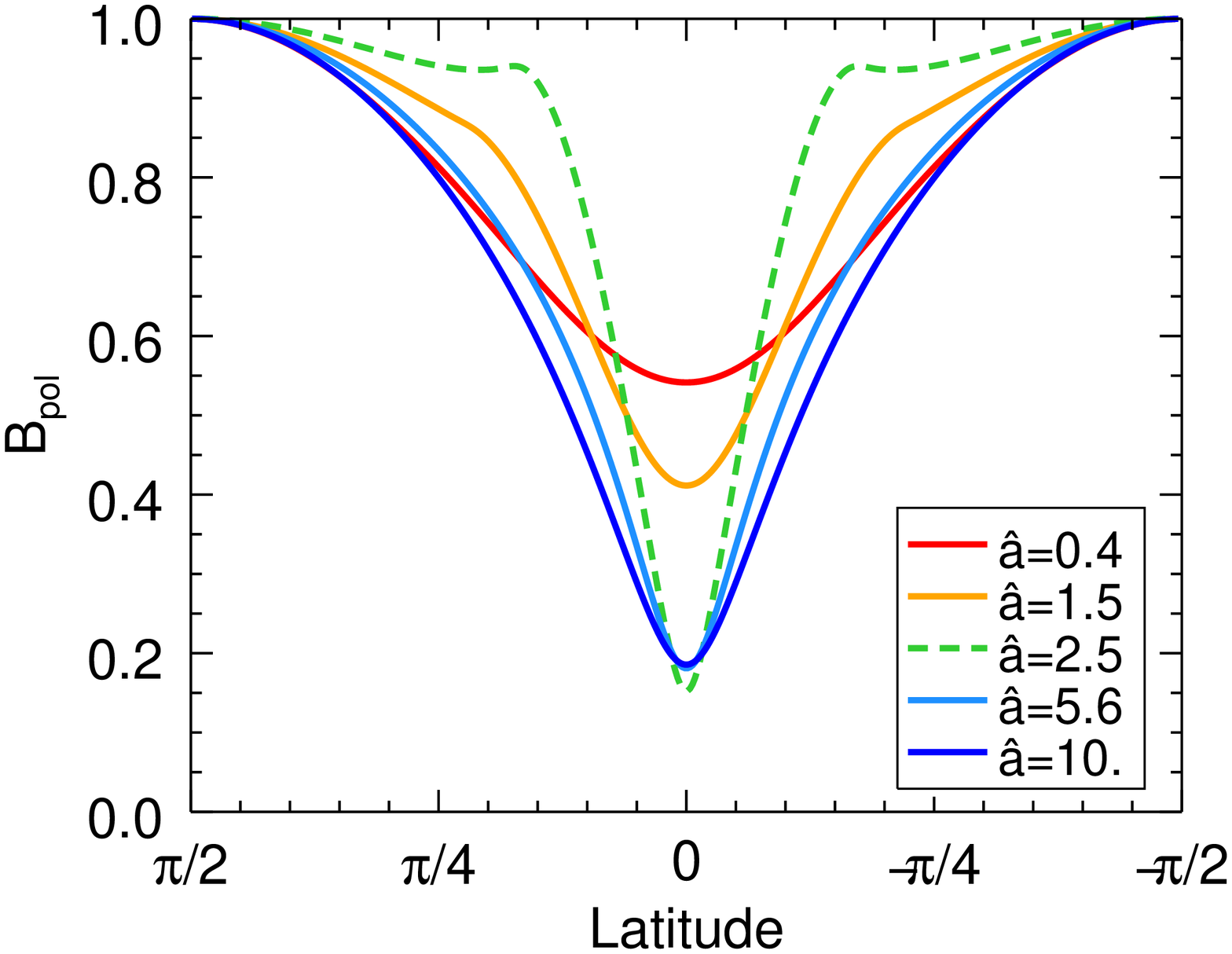}
	\includegraphics[width=.33\textwidth]{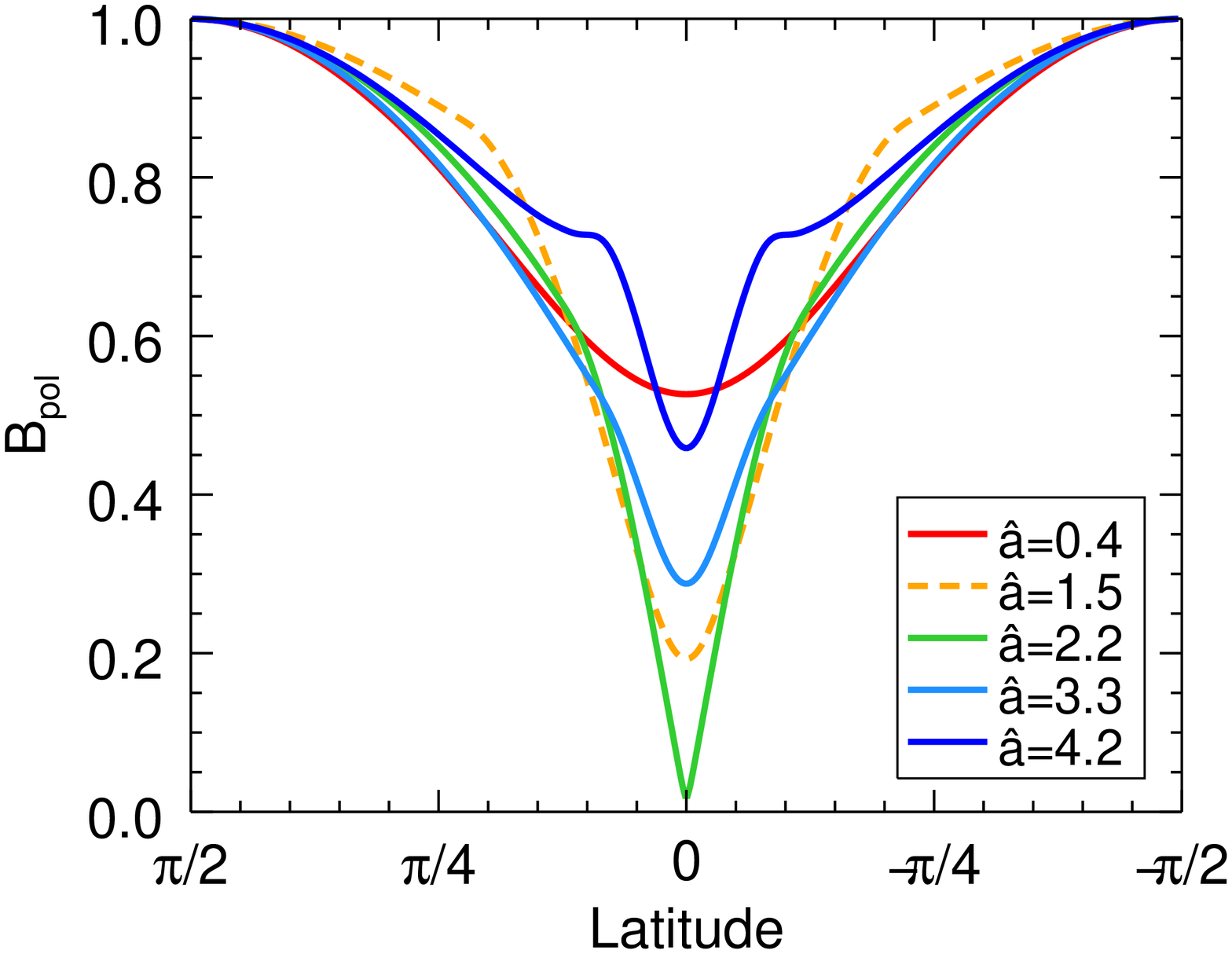}
	\includegraphics[width=.33\textwidth]{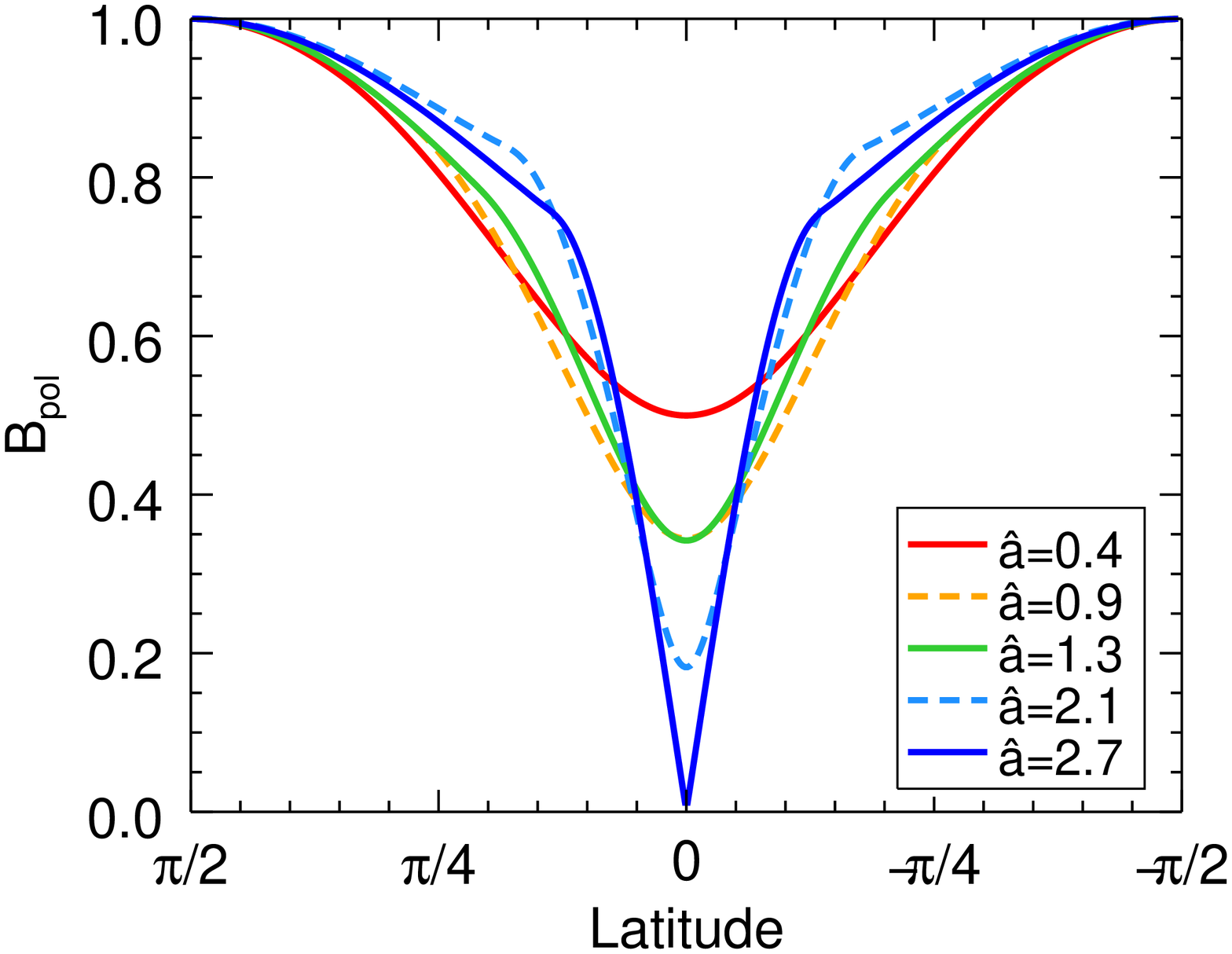}\\
	\includegraphics[width=.33\textwidth]{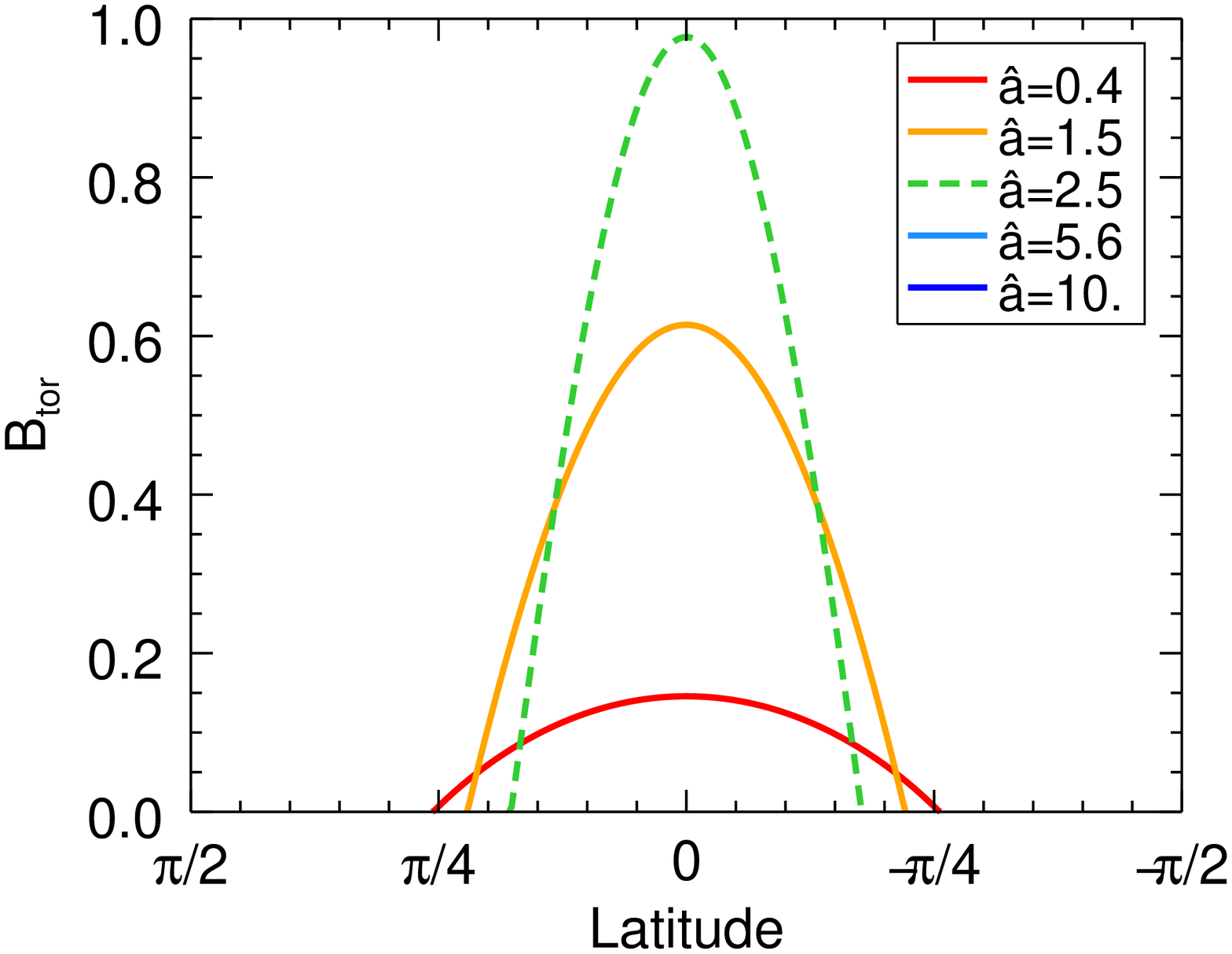}
	\includegraphics[width=.33\textwidth]{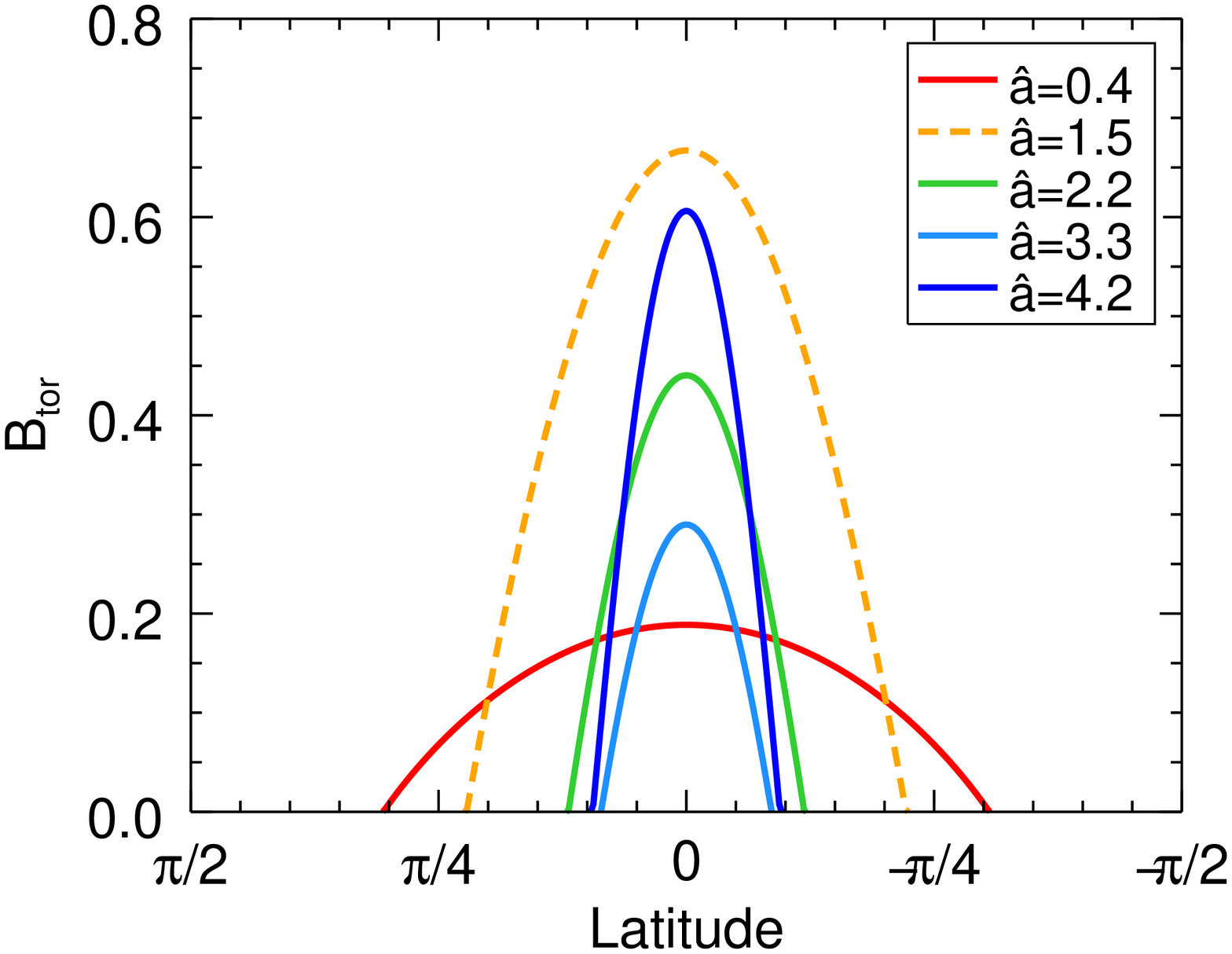}
	\includegraphics[width=.33\textwidth]{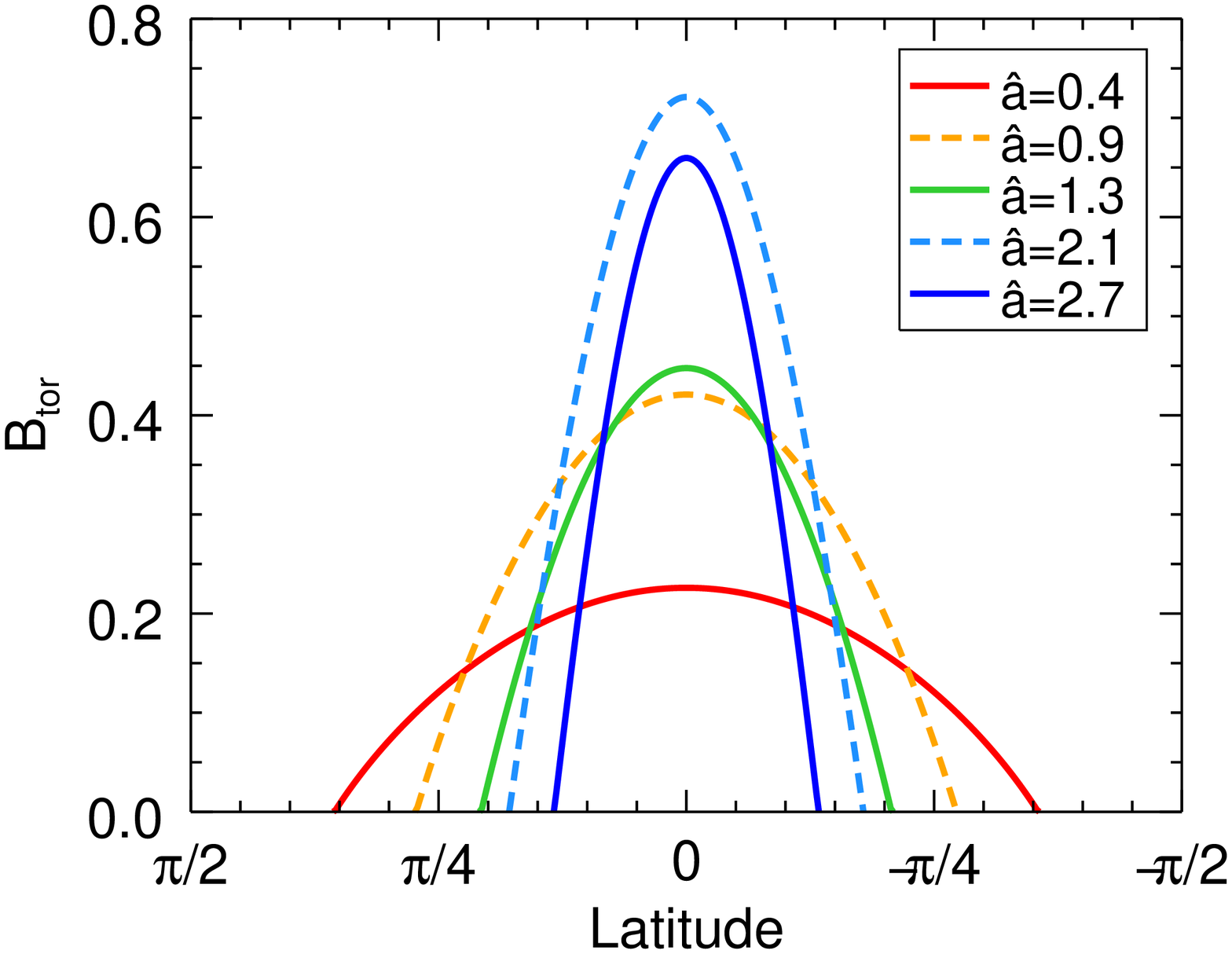}
	\caption{ Profiles of the strength (expressed in units of
	$B_{\rm pole}$) of the poloidal (top panel) and toroidal 
	(bottom panel) magnetic field on the stellar surface for 
	selected models ($\hat{a}$ is expressed in 
	units of $10^{-3}$)  along the equilibrium sequences 
	with $\lambda=2$ (left), $\lambda=3$ (centre) and  $\lambda=6$ 
	(right).	The dashed lines represent models where the value of 
	$\mathcal{H}_{\rm tor}/\mathcal{H}$ reaches a maximum.
              }
	\label{fig:Bsurf}
\end{figure*}

\begin{figure}
	\centering
    \includegraphics[width=.5\textwidth]{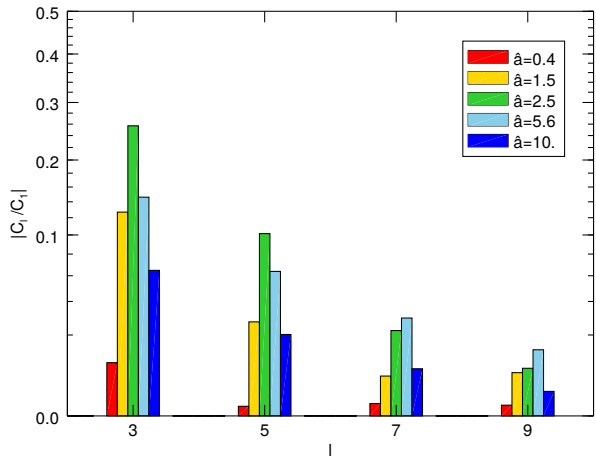}
	\caption{ Harmonic content $C_l/C_1$ of the magnetic field at 
	the	surface of the star in the case $\lambda=2$ for the selected models shown in
	the upper left panel of figure~\ref{fig:Bsurf}.
              }
	\label{fig:harmonic}
\end{figure}

In  figure~\ref{fig:Bsurf} we show the profiles of the poloidal and
toroidal components of the  magnetic 
field along the stellar surface. 
In the small $\hat{a}$ regime, the poloidal field
at the surface is essentially dipolar.
The toroidal magnetic field extends over a region $\pm 45^\circ$
in latitude around the equator, slightly bigger for larger values of $\lambda$. 
Trends are different depending if the structure evolves to a single peak or double
peak. For $\lambda  \le 2$ (single peak), as $\hat{a}$ increases, the
magnetic field becomes slightly higher in the polar region but
decreases substantially at the equator. As the peak moves outwards, so
does the poloidal neutral line (where the poloidal field vanishes).
This is the reason why the equatorial field drops. The magnetic field
at the surface becomes closer to a split monopole: the curvature of
magnetic field surfaces diminishes, the radial component becomes more
uniform, except very close to the equator. The portion of the surface
where $B_{\rm tor} \neq 0$ instead shrinks, and vanishes completely
once the twisted region gets out of the star. 
In figure~\ref{fig:harmonic} we show the harmonic content $C_l/C_1$ of the 
surface magnetic field for different  models, in the case $\lambda=2$. 
As expected the dipole term is always the dominant one. Multipole
terms become more important in correspondence to configurations
with the higher $\mathcal{H}_{\rm tor}/\mathcal{H}$, when the 
toroidal field is stronger and the neutral line is located just underneath the
stellar surface. Finally, as the twisted magnetospheric torus moves away from the star,
the multipolar content of the surface field drops.

For higher values of $\lambda \ge 3$ the appearance of multiple peaks
and disconnected magnetic regions leads to a more complex behaviour of
the magnetic field at the surface. In the polar regions at high
latitude $> 45^\circ$ the value of the poloidal field does not change
much with increasing $\hat{a}$. As  $\hat{a}$ increases the value of
the poloidal field at the equator drops, but in this case this is not
due to the neutral line moving outwards, but because an X-point
forms in the vicinity of the surface. Indeed, as can be seen in 
figures~\ref{fig:Bsurf} and \ref{fig:manymod}, in the $\lambda=3$ case the values of $\hat{a}$ for
which the poloidal field vanishes at the equator is the same at which
a second peak forms. For higher values  of $\hat{a}$, the equatorial
poloidal field rises again. The portion of the surface
where $B_{\rm tor} \neq 0$ shrinks again, though for these cases it never vanishes completely.
In all cases we find that the strength of the toroidal
component of the magnetic field at the surface tends to grow becoming
comparable to, or even exceeding, the strength of the poloidal one.

\begin{figure*}
	\centering
    \includegraphics[width=.33\textwidth]{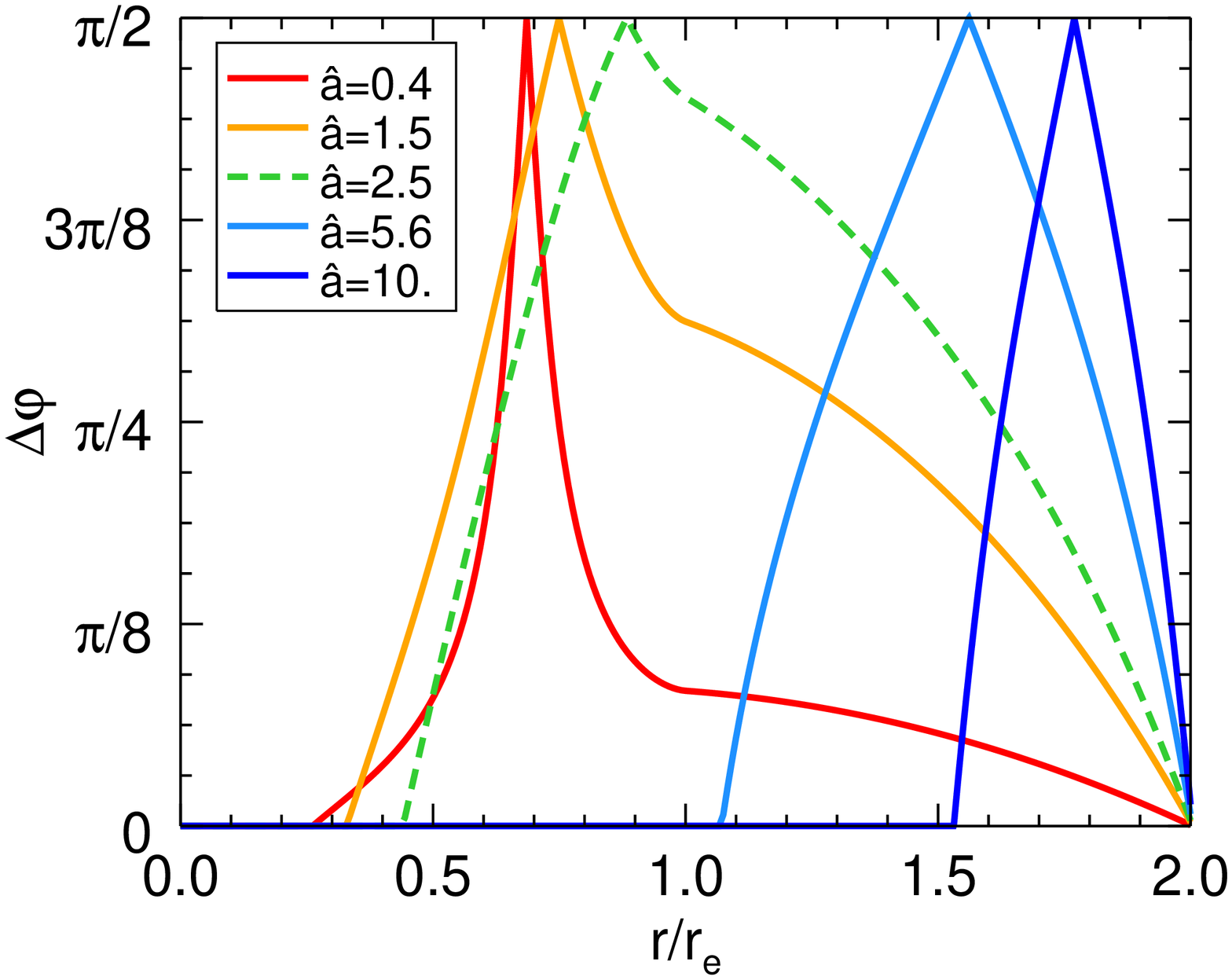}
	\includegraphics[width=.33\textwidth]{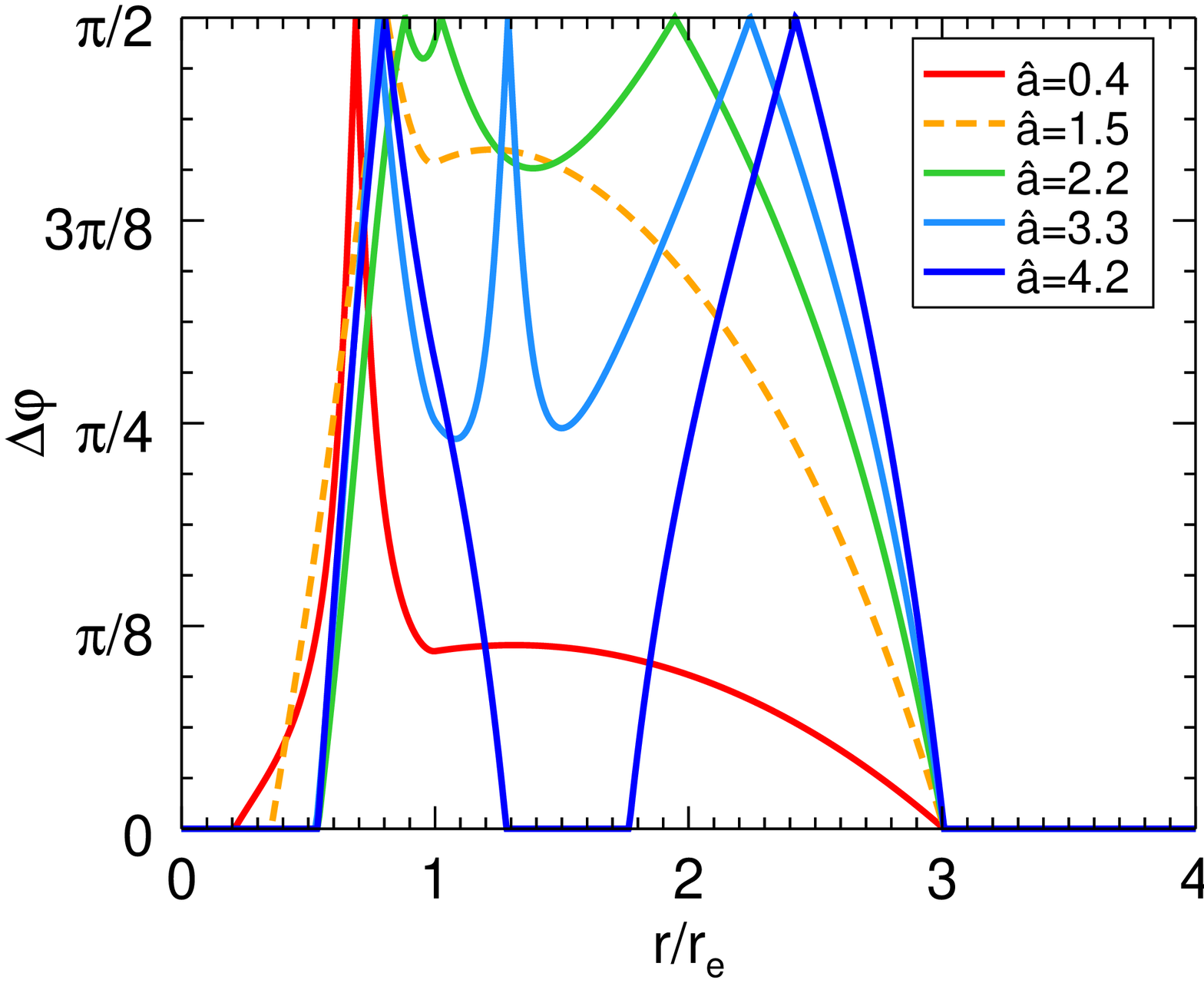}
	\includegraphics[width=.33\textwidth]{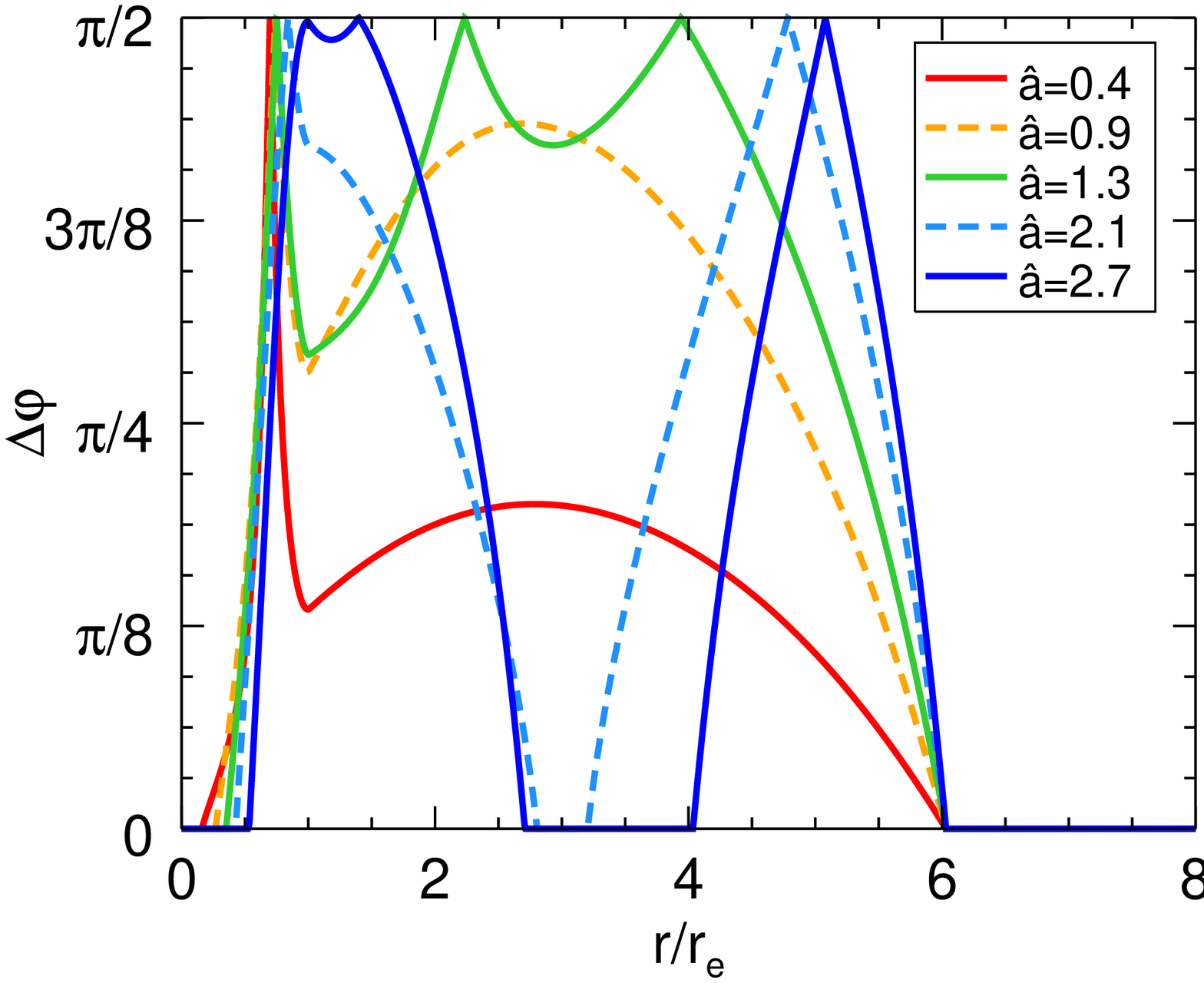}\\
	\includegraphics[width=.33\textwidth]{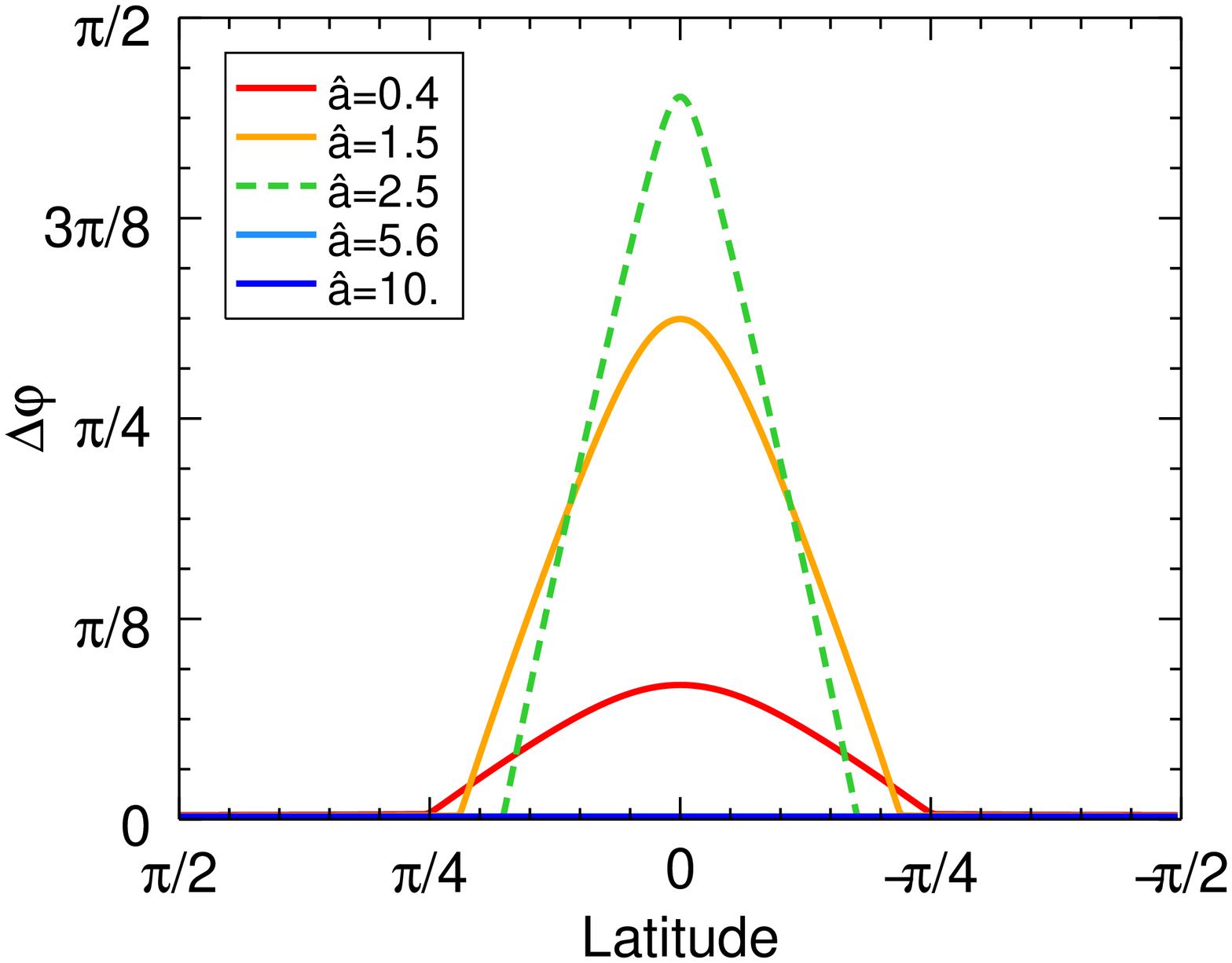}
	\includegraphics[width=.33\textwidth]{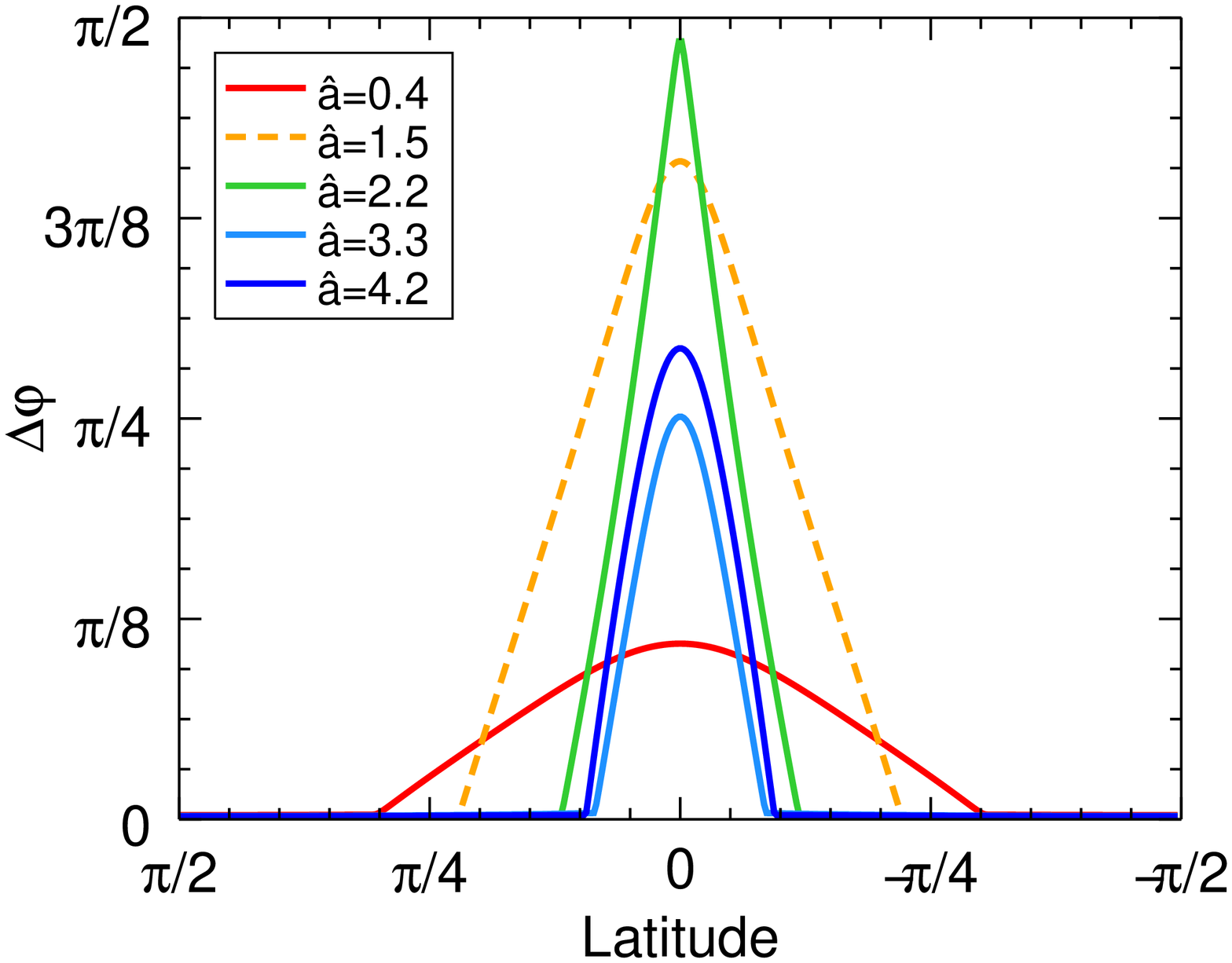}
	\includegraphics[width=.33\textwidth]{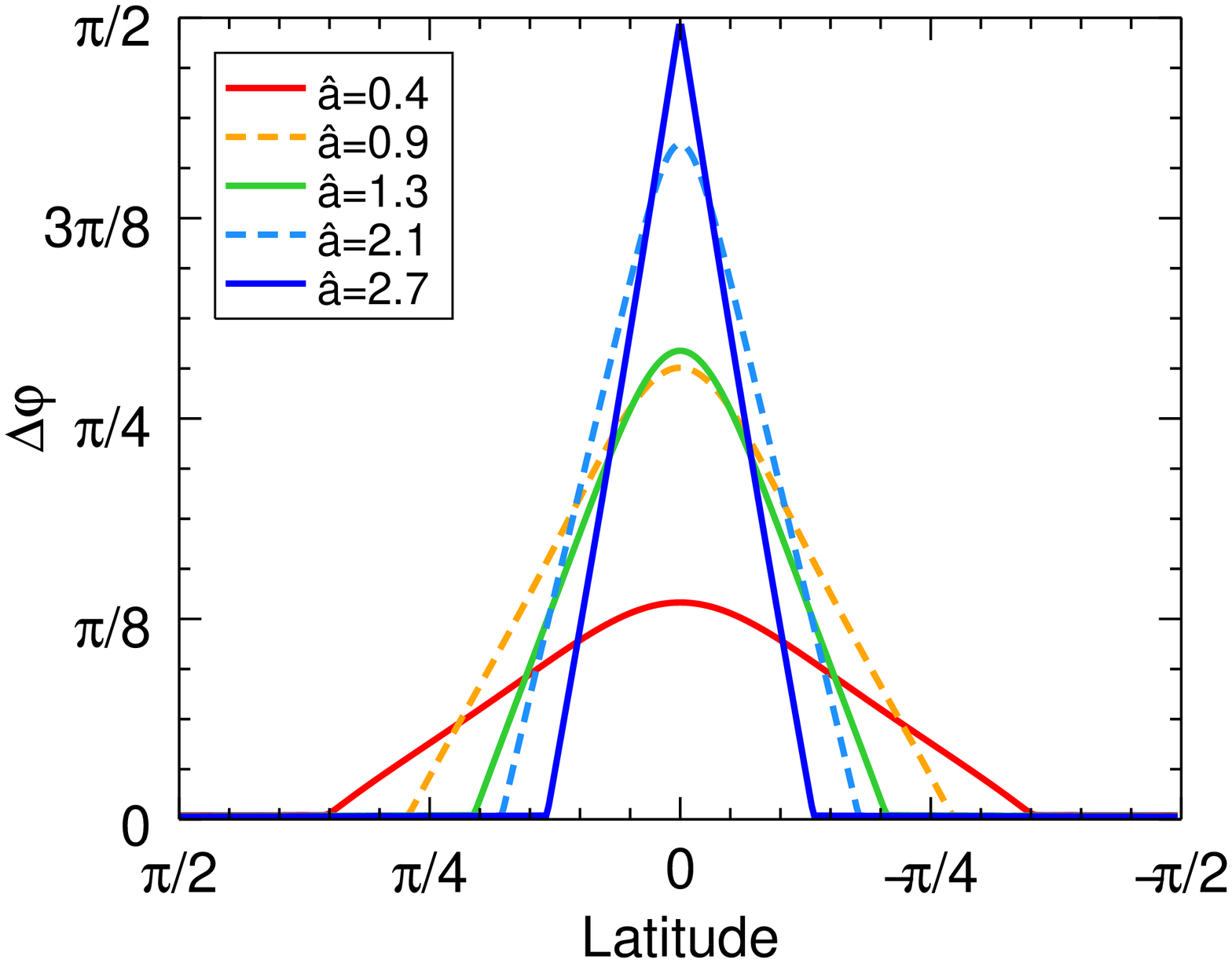}
	\caption{Profiles of the twist angle $\Delta \varphi$ at the
          equator as a function of distance (normalized to the stellar
          radius $r_e$) (upper panels), and as a function of latitude on the 	
			  stellar surface (bottom panels), for selected models 
			  ($\hat{a}$ is expressed in units of $10^{-3}$)  
			  along the equilibrium sequences with 
			  $\lambda=2$ (left), $\lambda=3$ (centre) and  $\lambda=6$ (right).
			  The dashed lines represent models with maximum value of 
			  $\mathcal{H}_{\rm tor}/\mathcal{H}$ along the same sequences.
              }
	\label{fig:deltaphi}
\end{figure*}
Another parameter that it is useful to show is the angle between
the magnetic field and the meridional plane, to which we refer as twist 
angle $\Delta \varphi$. For convenience it can be defined as the complementary 
of the angle that the magnetic field forms with respect to the azimuthal direction
\be
\Delta \varphi = \arcsin \left[ \left(B_\phi B^\phi /B_i B^i \right)^{1/2} \right].
\ee         
The value of $\Delta \varphi$ ranges from 0 to $\upi/2$, assuming
the latter when the magnetic field is purely azimuthal, 
and $0$  when the magnetic field lies in the meridional plane.
In figure~\ref{fig:deltaphi} we show this angle, measured at
the equator, as a function of radial distance $r$. 
Since our models are not rotating, all the magnetic field lines 
are eventually closed. The location where $\Delta \varphi = \upi/2 $
corresponds to the location of either the neutral line (O-point) or the X-point,
where the poloidal component of the magnetic field vanishes.
For cases with $\lambda \le 2$ the profile of $\Delta \varphi$ shows a unique
peak where the $\theta$-component of the poloidal field reverses sign
in the twisted region. Again we see that at higher values of $\hat{a}$ the toroidal 
field is completely outside the star.

As expected, in the case  $\lambda \ge 3$ the behaviour is more complex.
For the smallest value of $\hat{a}$ the trend of $\Delta \varphi$ 
resembles that of the analogous configuration at $\lambda=2$:
the twist is prominent in the vicinity of the neutral line
and extends outside the star remaining well below $\sim \upi/4$.
Moving at higher $\hat{a}$, the presence of 
two peaks in the toroidal magnetic field strength inside
the twisted region means that $\Delta \varphi$ reaches 
$\pi/2$ in three locations. In particular the first and
the third of those locations are always associated 
with O-points, the second one with an X-point.
At the higher value of $\hat{a}$, the formation of two detached
twisted regions is also evident. In this cases, however, the trend of 
$\Delta\varphi$ reveals only the position of the two O-point because 
the $\phi$-component of the magnetic field vanishes in correspondence 
with the X-point. Here we note that in the most extreme 
case at $\lambda=6$ the trend of $\Delta \varphi$ strengthen the hypothesis 
concerning the development of a third peak in the toroidal 
magnetic field strength. Indeed the second peak in $\Delta \varphi$
corresponds to an O-point and an unresolved X-point.

The bottom row of figure~\ref{fig:deltaphi} displays the profile
of $\Delta \varphi$ along the stellar surface.
As pointed out before, in the limit of small $\hat{a}$ the twist at the
surface increases. However, for higher values of $\hat{a}$ the trend
is not uniform, depending on the formation of a second peak, and the
related location of the X-point. 
\begin{figure*}
	\centering
	\includegraphics[width=.45\textwidth]{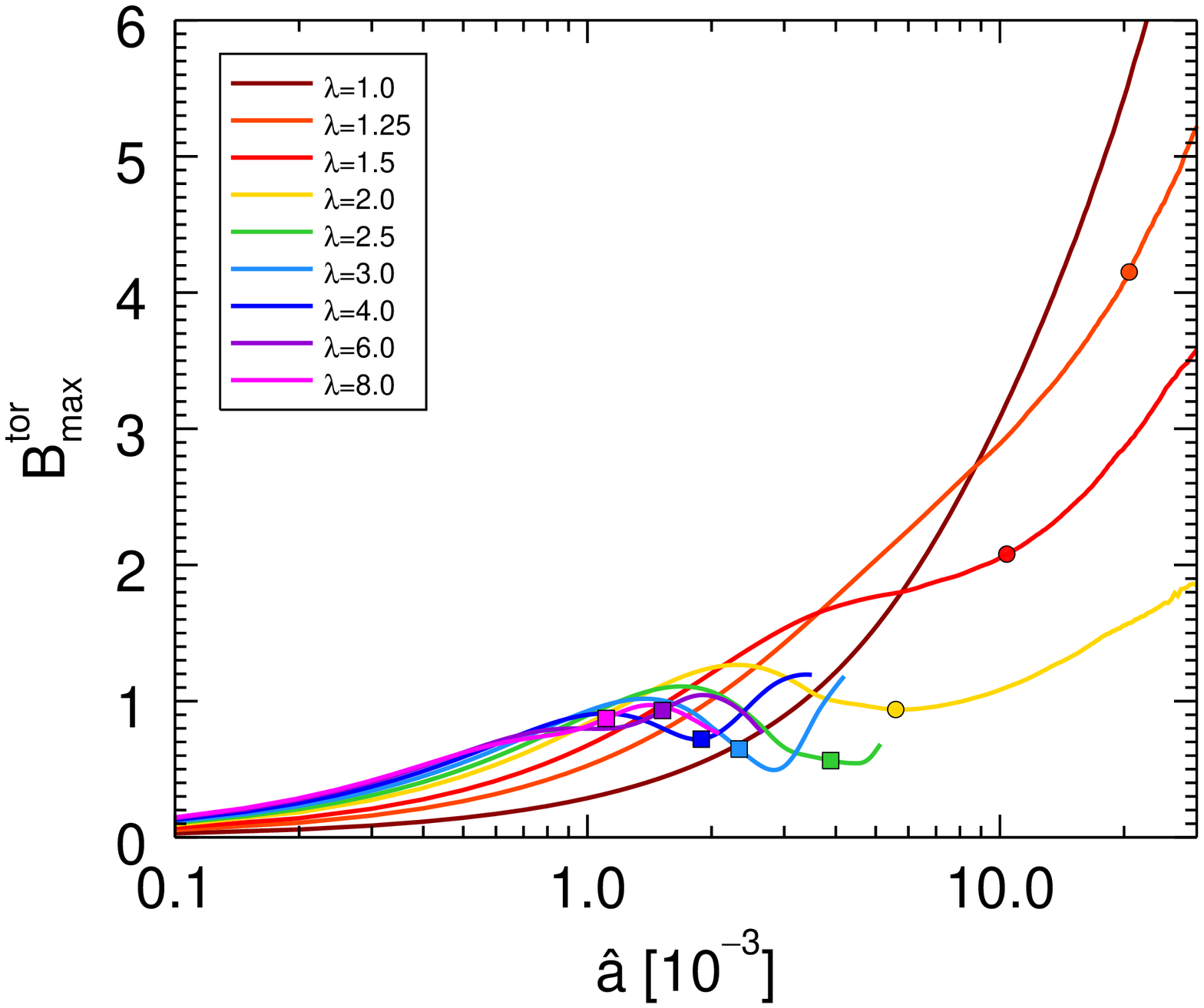}
	\includegraphics[width=.45\textwidth]{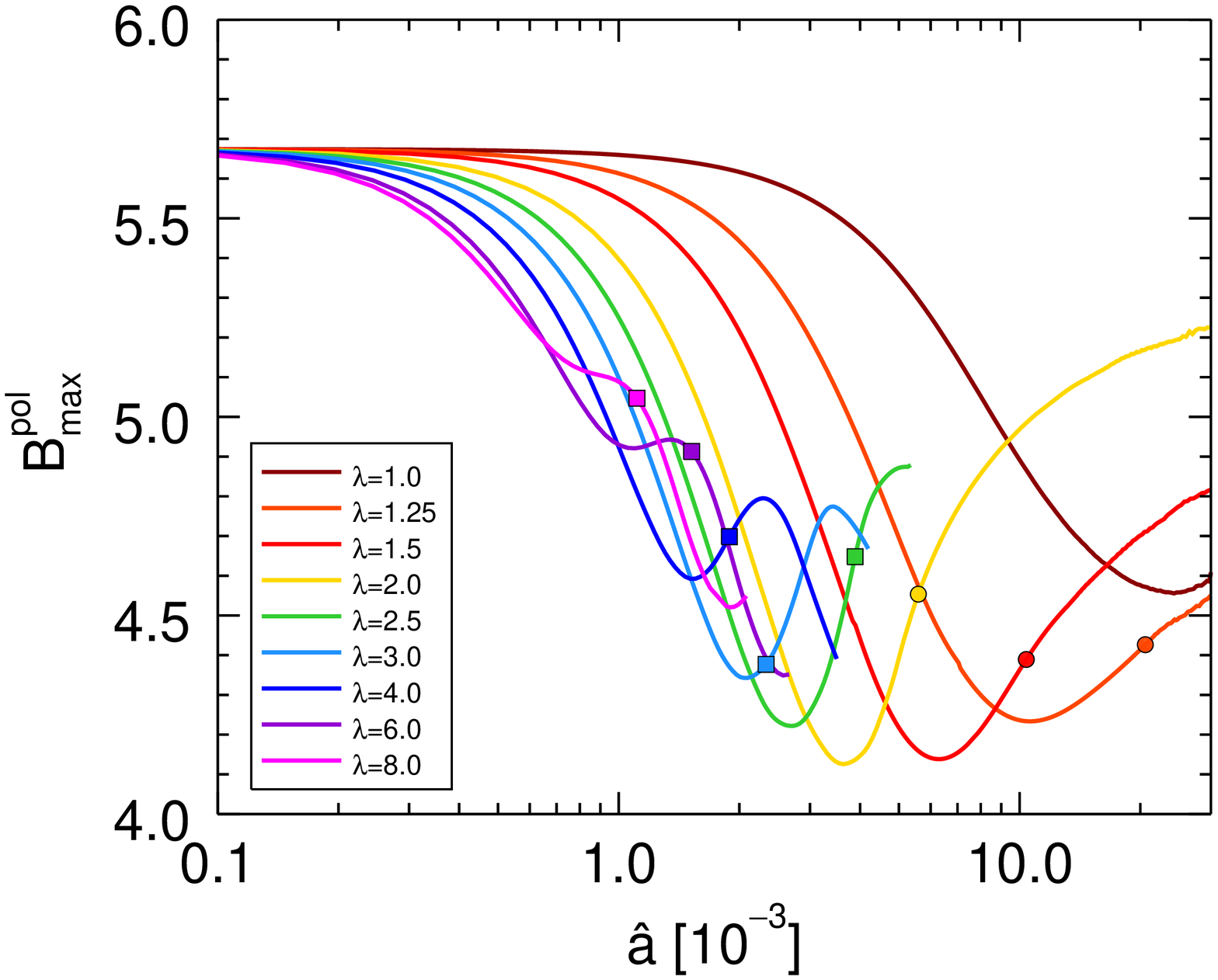}
	\caption{Maximum value of the toroidal (left) poloidal (right) magnetic field
	along equilibrium sequences with different values of $\lambda$.
	Dots indicate configurations where the toroidal magnetic field
        component is completely
	outside the star. Squares indicate configurations which show two maxima of the
	toroidal magnetic field.
              }
	\label{fig:bmax}
\end{figure*}

\begin{figure*}
	\centering
	\includegraphics[width=.33\textwidth]{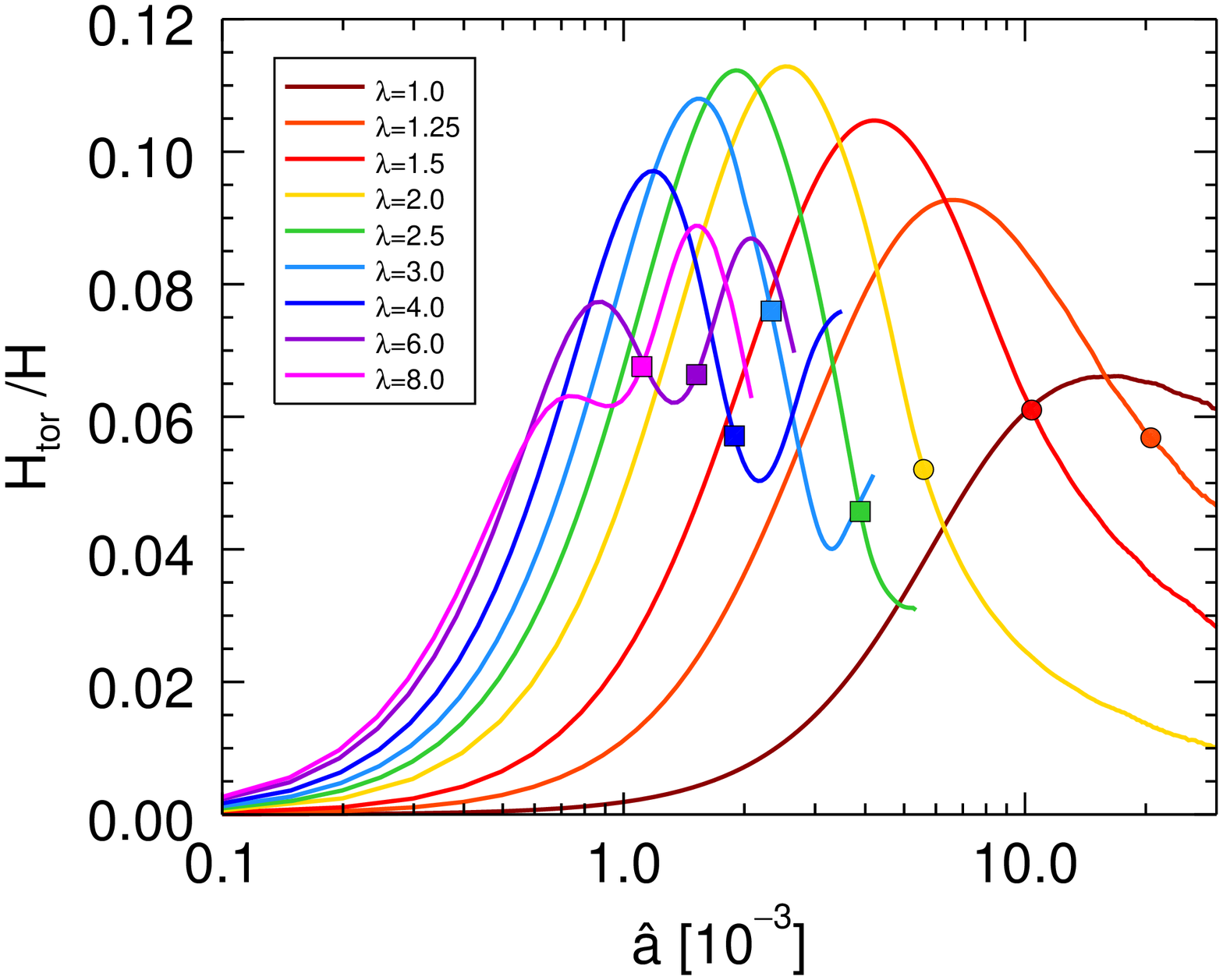} 
	\includegraphics[width=.33\textwidth]{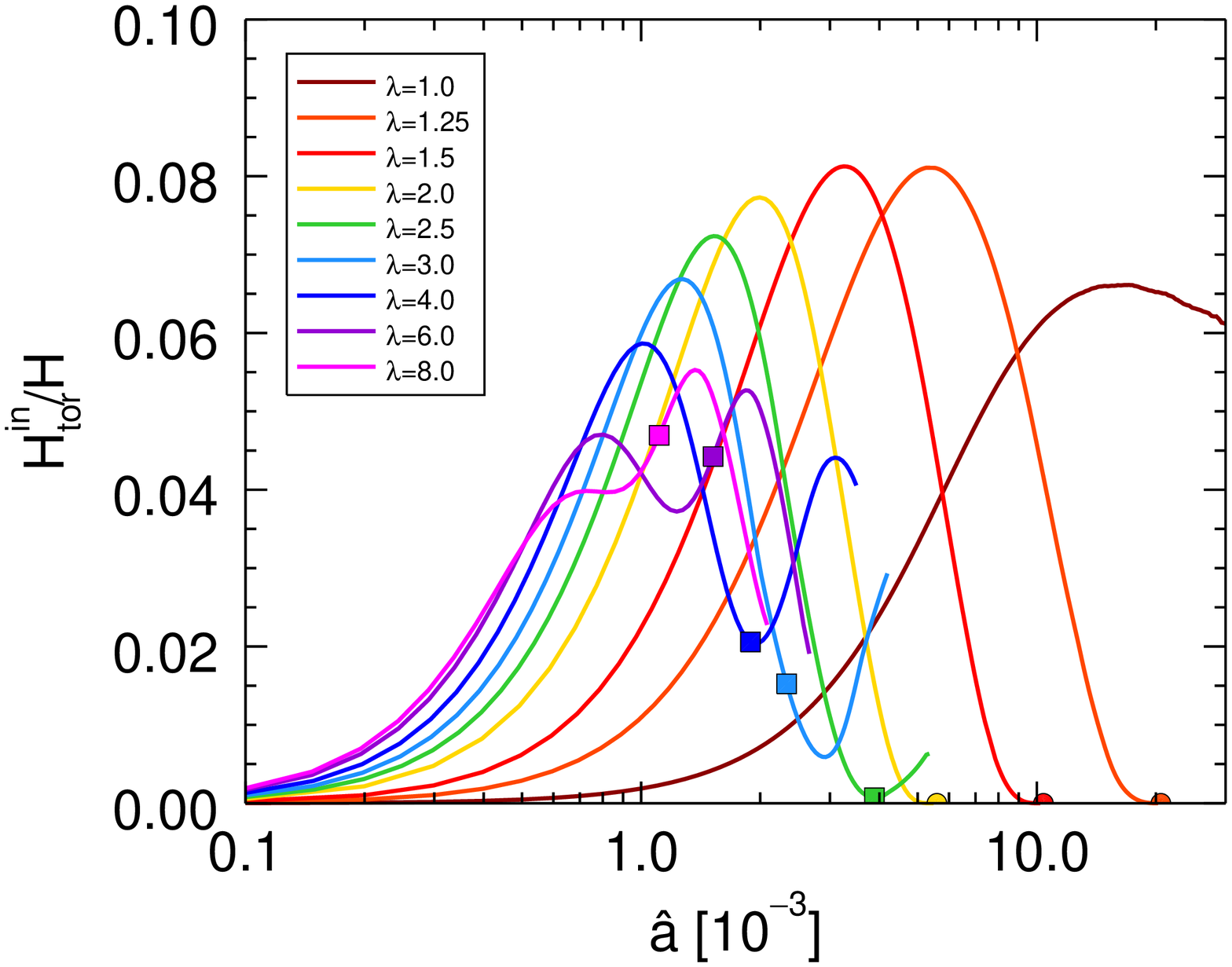}
	\includegraphics[width=.33\textwidth]{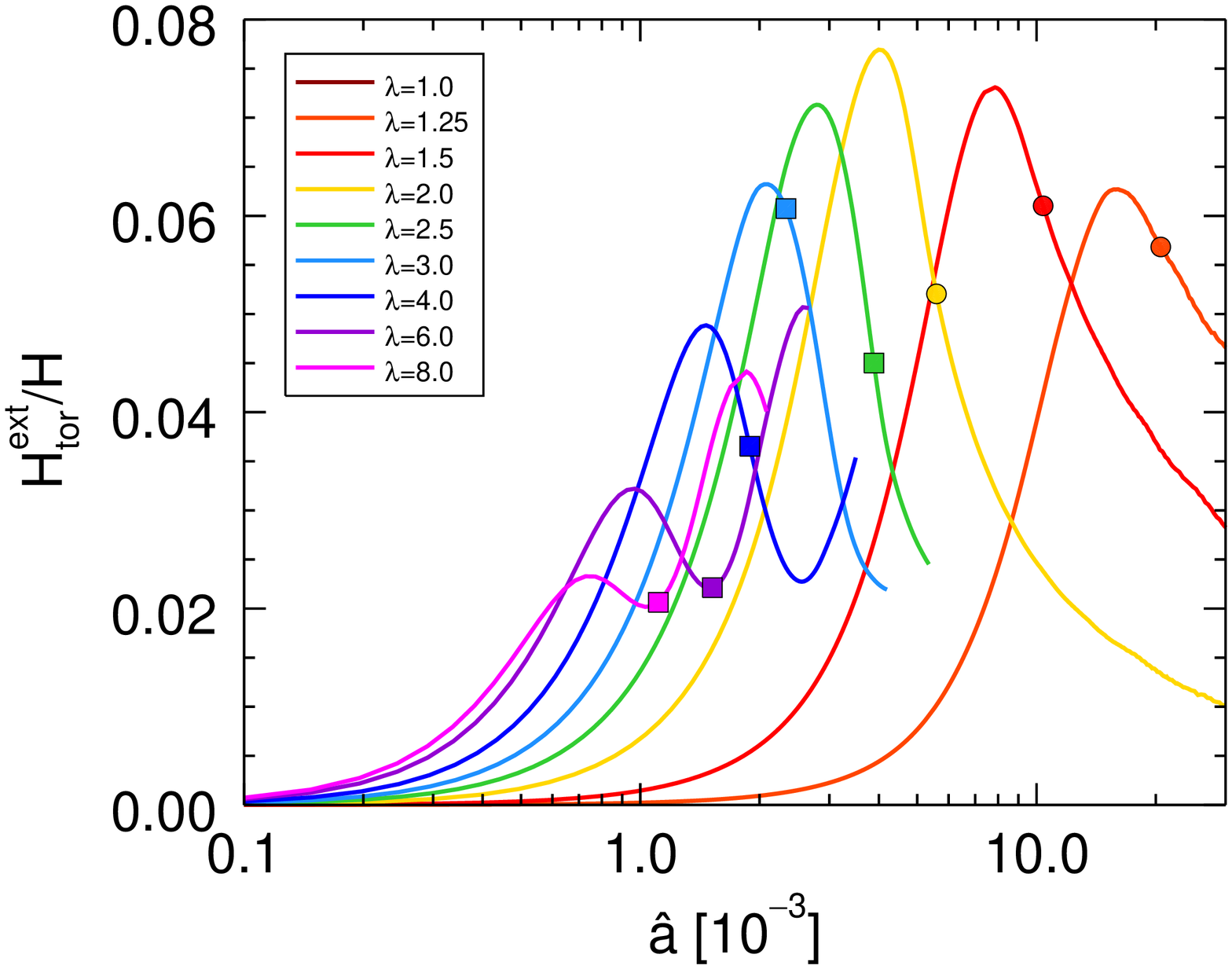}\\
	\caption{Profiles of the ratio of total magnetic energy of the
         toroidal component $\mathcal{H}_{\rm tor}$,
	     external one $\mathcal{H}_{\rm tor}^{\rm ext}$  and internal one
         $\mathcal{H}_{\rm tor}^{\rm in}$, with respect to the total
         magnetic energy $\mathcal{H}$. Curves show sequences
	     as a function of $\hat{a}$ for various values of $\lambda$.
		 Dots and squares as in figure \ref{fig:bmax}. }
	\label{fig:HoH}
\end{figure*}

In figure~\ref{fig:bmax} we plot the maximum value of the strength of
the toroidal magnetic field  $B_{\rm tor}^{\rm max}$, and the poloidal
one $B_{\rm pol}^{\rm max}$, for different values of $\lambda$, as a
function of $\hat{a}$. In all our models the poloidal field reaches
its maximum at the centre. Initially, in the small $\hat{a}$ regime,
$B_{\rm tor}^{\rm max}$ grows, while $B_{\rm pol}^{\rm max}$ decreases. 
This happens because the strength of the magnetic field at 
the pole is always kept fixed in all models. 
As one enhances the contribution to the total current by
increasing $\mathcal{I}$, one must decrease the contribution from
$\mathcal{M}$, causing a drop in the strength of the field at the
centre of the star. This effect depends also on the location of the
current, as  this term moves to larger radii the poloidal field begins to grow again.
Configurations with $\lambda \ge 3$ show several inversions of this
trend, which again are a manifestation of the change in the field topology. 

All the equilibrium models we obtain are energetically dominated by the poloidal 
magnetic field. This was found to apply also for models where the
twist is fully confined with the star (PBD14; \citealt{Lander_Jones09a,Ciolfi_Ferrari+09a}).
In figure~\ref{fig:HoH} we show the same equilibrium sequences in terms
of the ratio of magnetic energy of the toroidal magnetic field
$\mathcal{H}_{\rm tor}$ over the total  magnetic energy $\mathcal{H}$. 
Generally the magnetic energy ratio
initially grows with $\hat{a}$ reaching a first maximum that
corresponds to a configuration still characterized by a single peak (see
the first rows of fig \ref{fig:maxratio}). Again the trend for higher
values of $\hat{a}$  depends on the value of $\lambda$.
While sequences with $\lambda \leq 2$ show a decreasing monotonic trend, sequences
with $ \lambda \ge 3$ reach a minimum and then the magnetic energy ratio begin to 
grow again. For configurations with $ \lambda \gtrsim 6$ we could
reach a second local maximum. It is possible in principle that other maxima
and minima could be reaches at higher values of $\hat{a}$, but we
could not compute those model. The magnetic energy is an integrated
quantity, as such it also depends on the size of the twisted
region. The formation of an X-point, followed by the formation of
two detached magnetic twisted domains, is associated to a decrease of
the net volume taken by the toroidal field, and to the drop of
$\mathcal{H}_{\rm tor}$  after the first maximum.

In figure~\ref{fig:HoH} we also compare the toroidal magnetic energy
confined inside $\mathcal{H}_{\rm tor}^{\rm in}$ 
and outside $\mathcal{H}_{\rm tor}^{\rm ext}$ the star. The two are in
general comparable except for cases with $\lambda \le 2$ where the
interior toroidal field vanishes at high $\hat{a}$. 
Note also that the ratio $\mathcal{H}_{\rm tor}/\mathcal{H}$ is at
most 8-10\%. The net poloidal and toroidal currents follow a similar behaviour.

\begin{figure*}
	\centering
	\includegraphics[width=.45\textwidth]{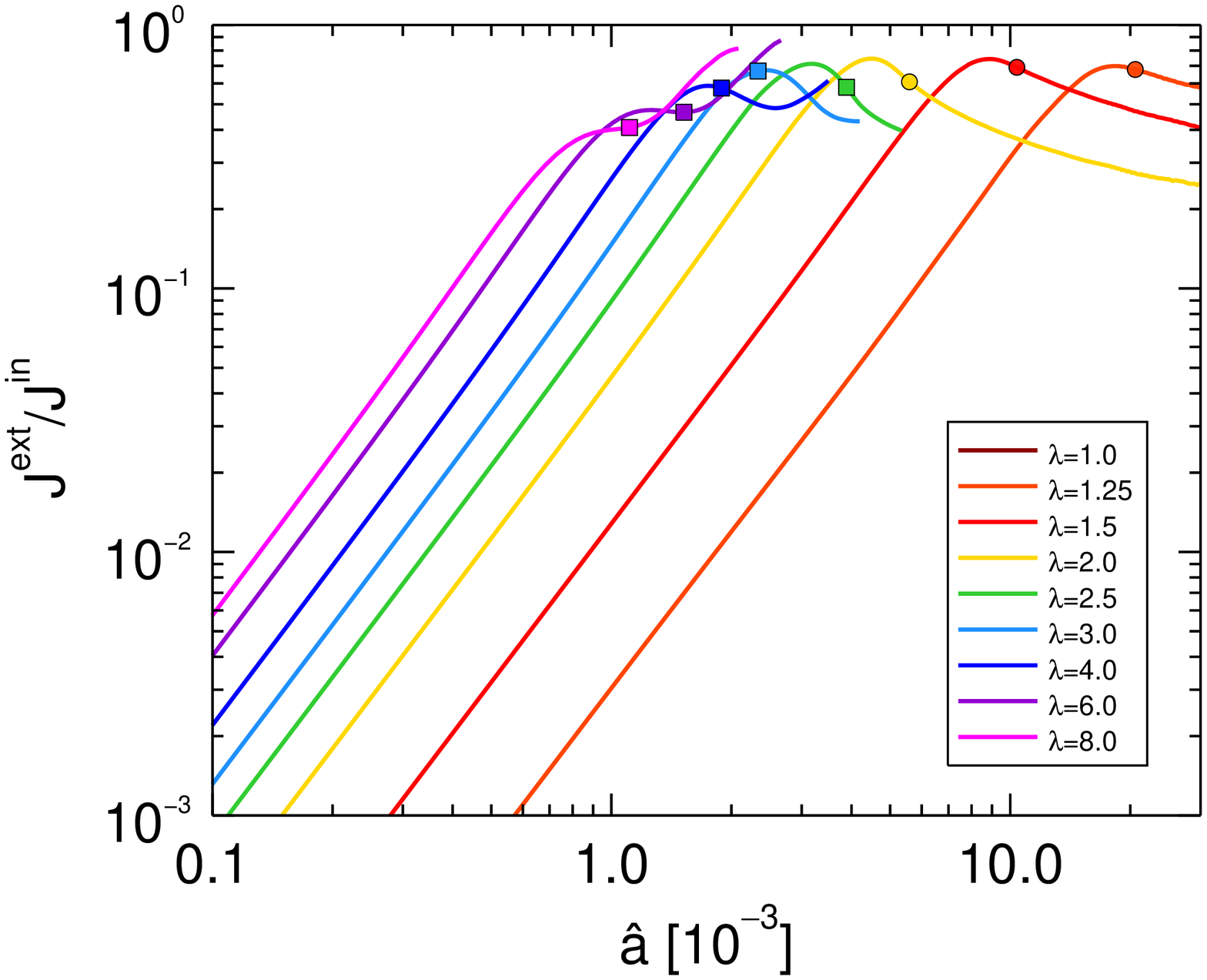} 
	\includegraphics[width=.45\textwidth]{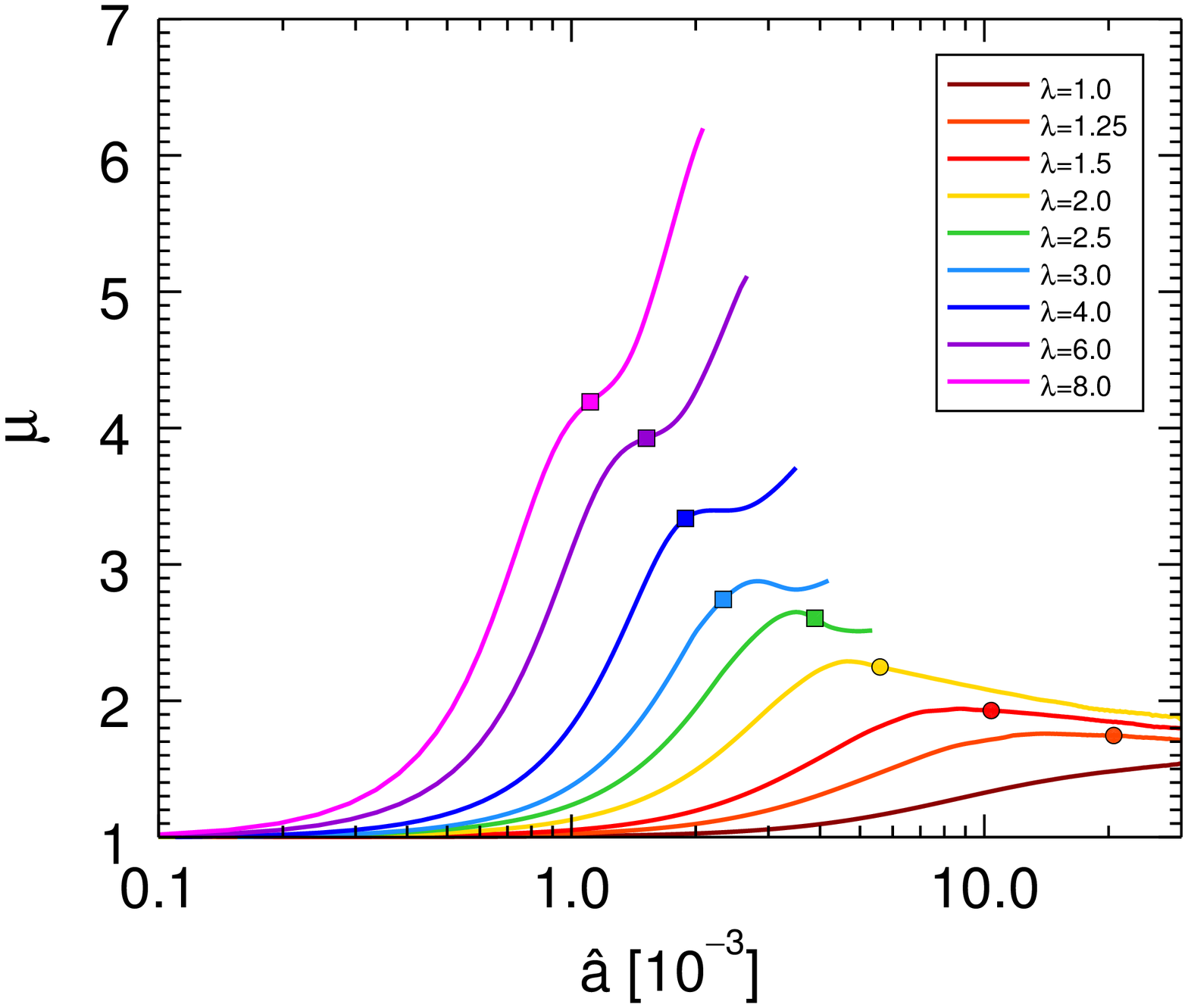}
	\caption{Left panel: ratio between the toroidal current density
    outside the star $\mathcal{J}^{\rm ext}$ and
	the toroidal current density inside $\mathcal{J}^{\rm in}$, as
    a function of $\hat{a}$ along
	equilibrium sequences with different value of $\lambda$.
	Right: magnetic dipole moment $\mu$ along the same sequences
	expressed in units of the magnetic dipole moment of the 
	fiducial configuration
	$\mu=1.114 \times 10^{32} \, \mathrm{erg}/\mathrm{G}$.
	Dots and squares as in figure \ref{fig:bmax}. 
              }
	\label{fig:joj}
\end{figure*}

Finally in figure~\ref{fig:joj} we show the variation of the magnetic dipole moment $\mu$
as a function of $\hat{a}$. We see that the magnetic dipole, for fixed values of $\hat{a}$,
grows with $\lambda$. This is because the total integrated toroidal current,  
defined as
\be
\mathcal{J}= \int \sqrt{J^{\phi}J_{\phi}} \psi^{6} r^2 \sin \theta dr d\theta d\phi
\ee
is bigger for larger magnetospheres with higher values of $\lambda$.
Very interestingly, for large values of $\lambda$ the net magnetic
dipole moment can be even 4-5 times higher, given the same strength of
the field at the pole. External currents contribute to the net dipole
without affecting too much the strength of the magnetic field at the
surface. This is a known property of twisted magnetospheres \citep{Thompson_Lyutikov+02a}.
 
For all the configurations computed here the internal linear
current  $\mathcal{J}^{\rm in}$ is always greater than the external
one $\mathcal{J}^{\rm ext}$, reaching similar values only for
configuration where the energy ratio reaches a maximum.
 At first, as expected, the external current, due only to the term
 $\mathcal{I}$, grows linearly with $\hat{a}^2$,
 while the internal one dominated by $\mathcal{M}$ remains more or
 less constant.
For higher values of $\hat{a}$ the ratio decreases exactly for the same
volumetric effect that was discussed for  the
trend of $\mathcal{H}_{\rm tor}/\mathcal{H}$.

\subsection{Models with $\zeta=1$}

The toroidal magnetization index $\zeta$ controls the shape of the
current distribution inside the torus-like region of the twisted 
field. With respect to the $\zeta=0$ case, choosing higher value for $\zeta$ entails
stronger currents mostly concentrated in the proximity of the neutral line.
In this section we will consider the $\zeta=1$ case in order to show which are the possible
qualitative and quantitative differences that can arise if a different value of $\zeta$ is chosen.

\begin{figure*}
	\centering
	\includegraphics[width=.33\textwidth]{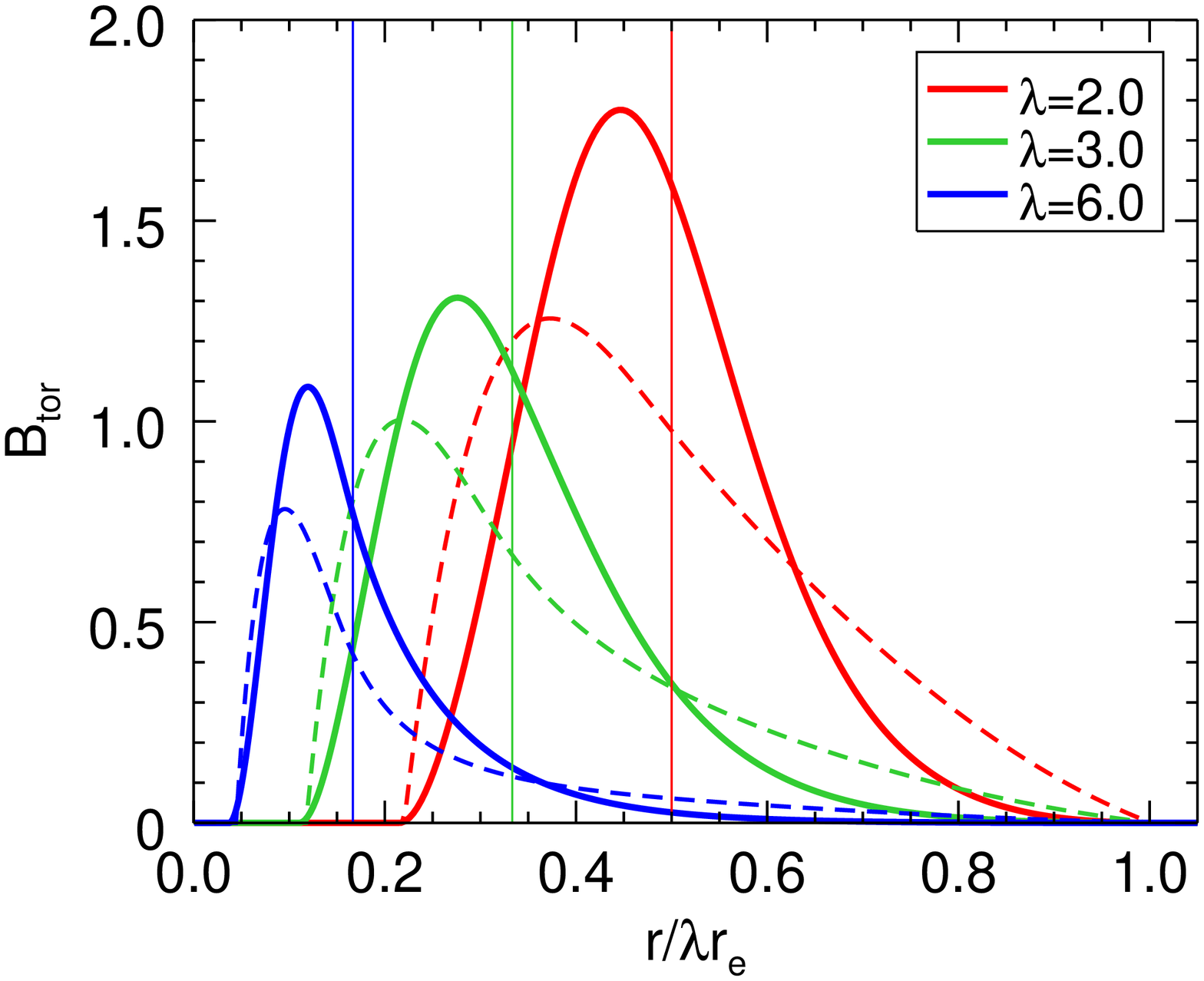}
	\includegraphics[width=.33\textwidth]{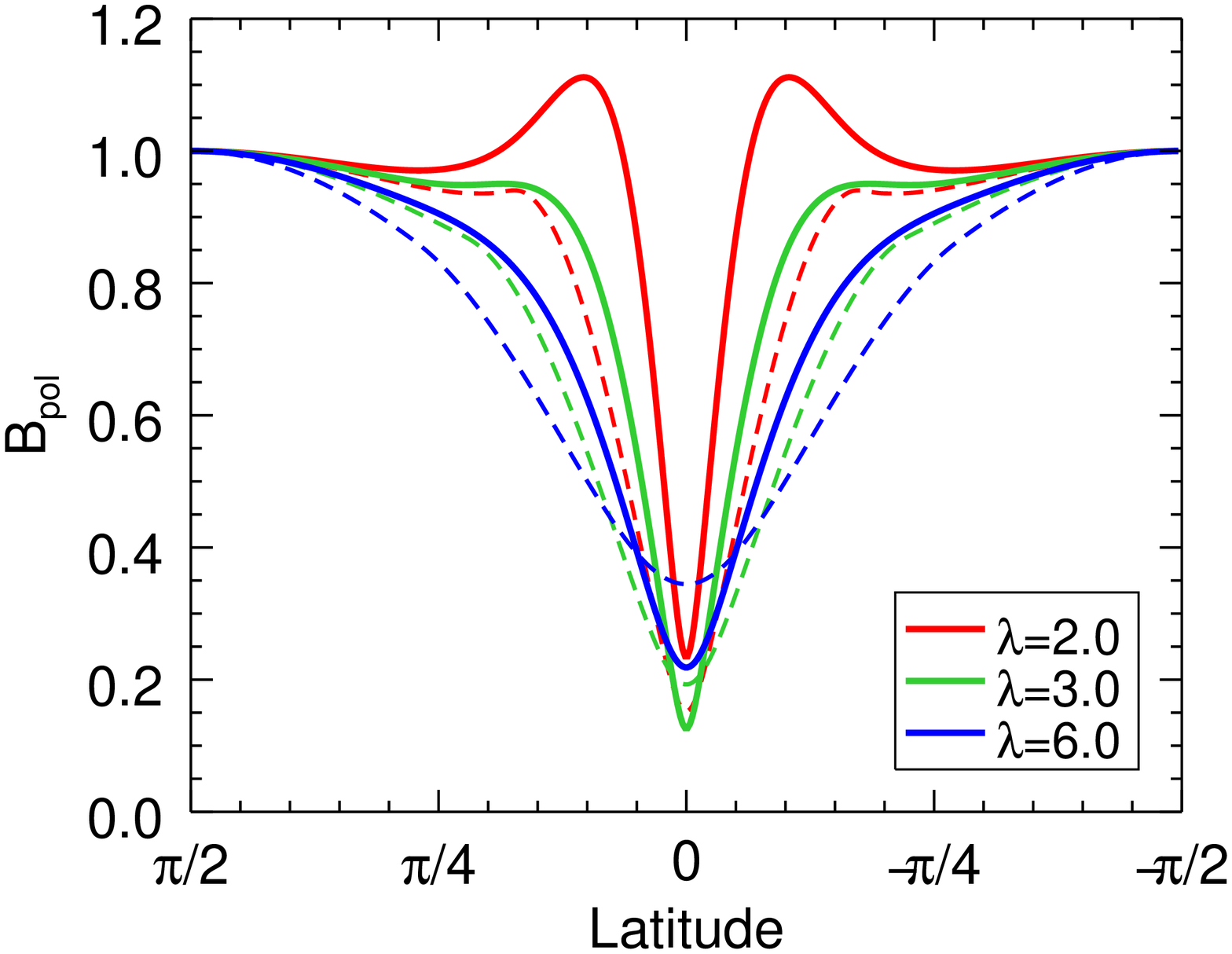}
	\includegraphics[width=.33\textwidth]{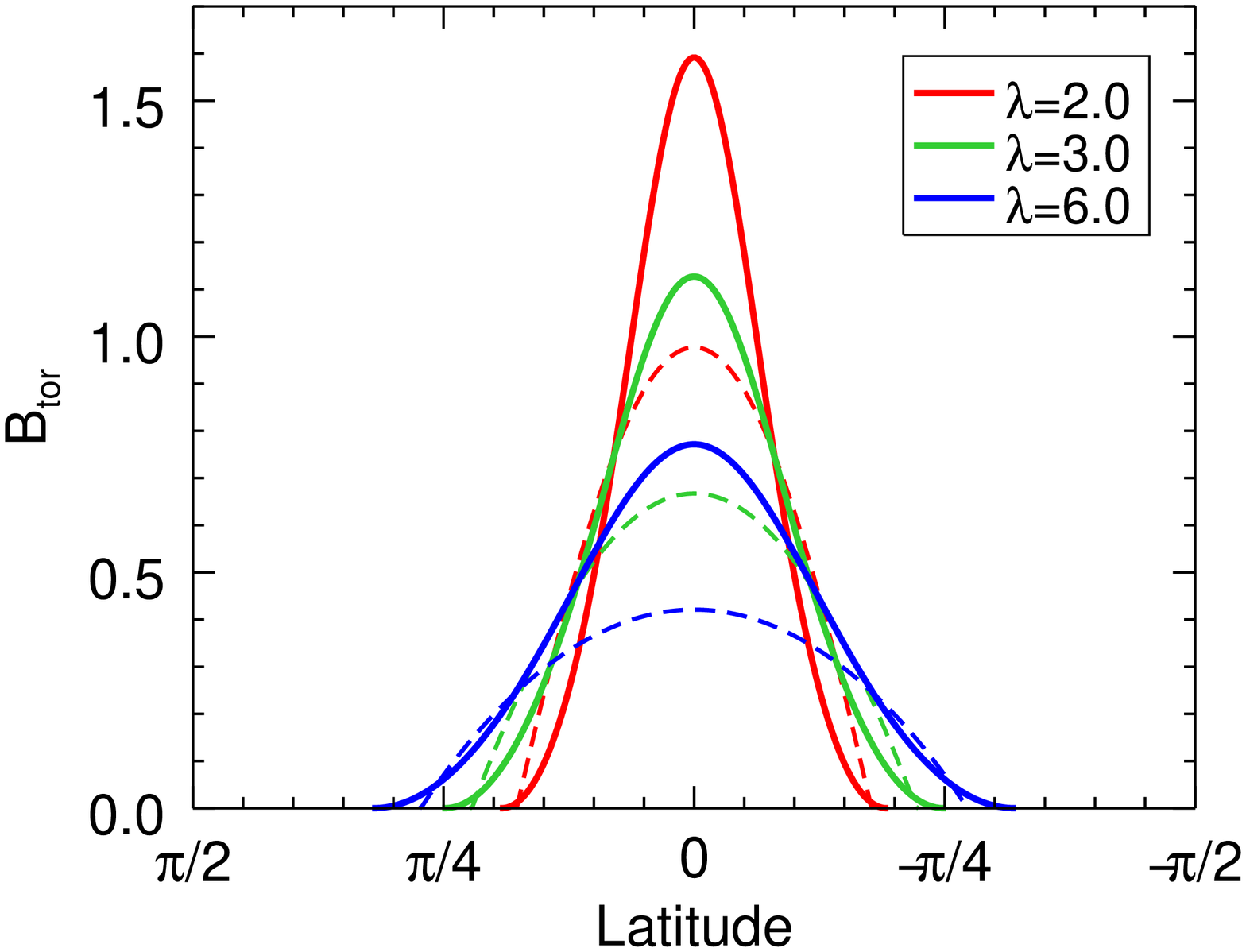}
	\caption{ The left panel shows the profile of the toroidal magnetic field strength 
	          (in units of $B_{\rm pole}$) for the configuration with the first maximum value of the 
	          magnetic energy ratio $\mathcal{H}_{\rm tor}/\mathcal{H}$ along sequences with fixed 
	          $\lambda=$2,3 and 6. Solid lines refer to
                  configurations with $\zeta=1$, while dashed ones
	          refer to the equivalent configurations with $\zeta=0$. The thin vertical lines
	          indicate the location of the stellar surface for each $\lambda$.
	          The remaining panels show, for the same configurations, the profile of the magnetic 
	          field strength, evaluated at the surface of the star, for the poloidal (center) and 
	          the toroidal component (right) as a function of  latitude.
	          }
	\label{fig:bprofZ1}
\end{figure*}

In order to compare the results with those at $\zeta=0$, let us focus
to those configurations where  $\mathcal{H}_{\rm
  tor}/\mathcal{H}$ is maximal.
In figure~\ref{fig:bprofZ1} we show the strength of the magnetic
field, both at the surface and along the equator, in the cases $\lambda=2,3$ and 6, compared with that of the equivalent 
configurations at $\zeta=0$.
In the $\zeta=1$ case the toroidal magnetic field reaches higher
values than in the $\zeta=0$
case. However, even though the geometry and shape of the twisted region remains almost
the same, the distribution of the magnetic field is more concentrated around
the peak and the magnetic field decays more rapidly to zero in the magnetosphere.

Looking at the distribution of the poloidal and toroidal field
at the surface of the star (central and right panel in 
figure~\ref{fig:bprofZ1}) it is evident that the multipolar terms
of the  magnetic field become more important in the
$\zeta=1$ cases: while the strength of the poloidal magnetic field
at the equator decreases marginally, it increases in the neighbouring 
region where it can also exceed the value of $B_{\rm pole}$ within a 
wedge of about $\pm \upi/4$ around the equator. While the portion
of the surface where $B_{\rm tor} \neq 0$ remains approximatively the same
the toroidal field is now more concentrated around the equator where its
strength can be a factor $\sim 2$ higher than for $\zeta=0$.

\begin{figure*}
	\centering
	\includegraphics[width=.45\textwidth]{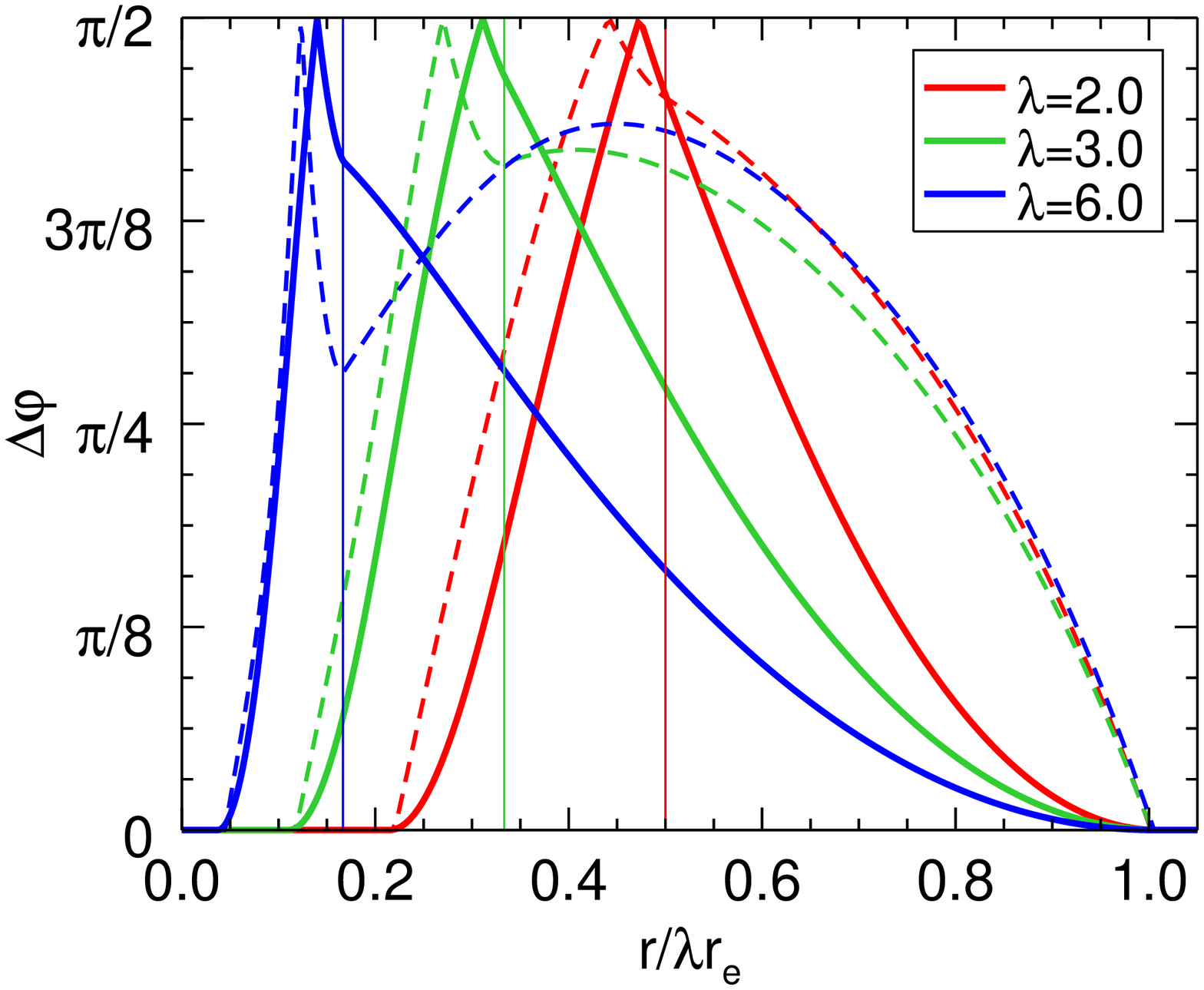}
	\includegraphics[width=.45\textwidth]{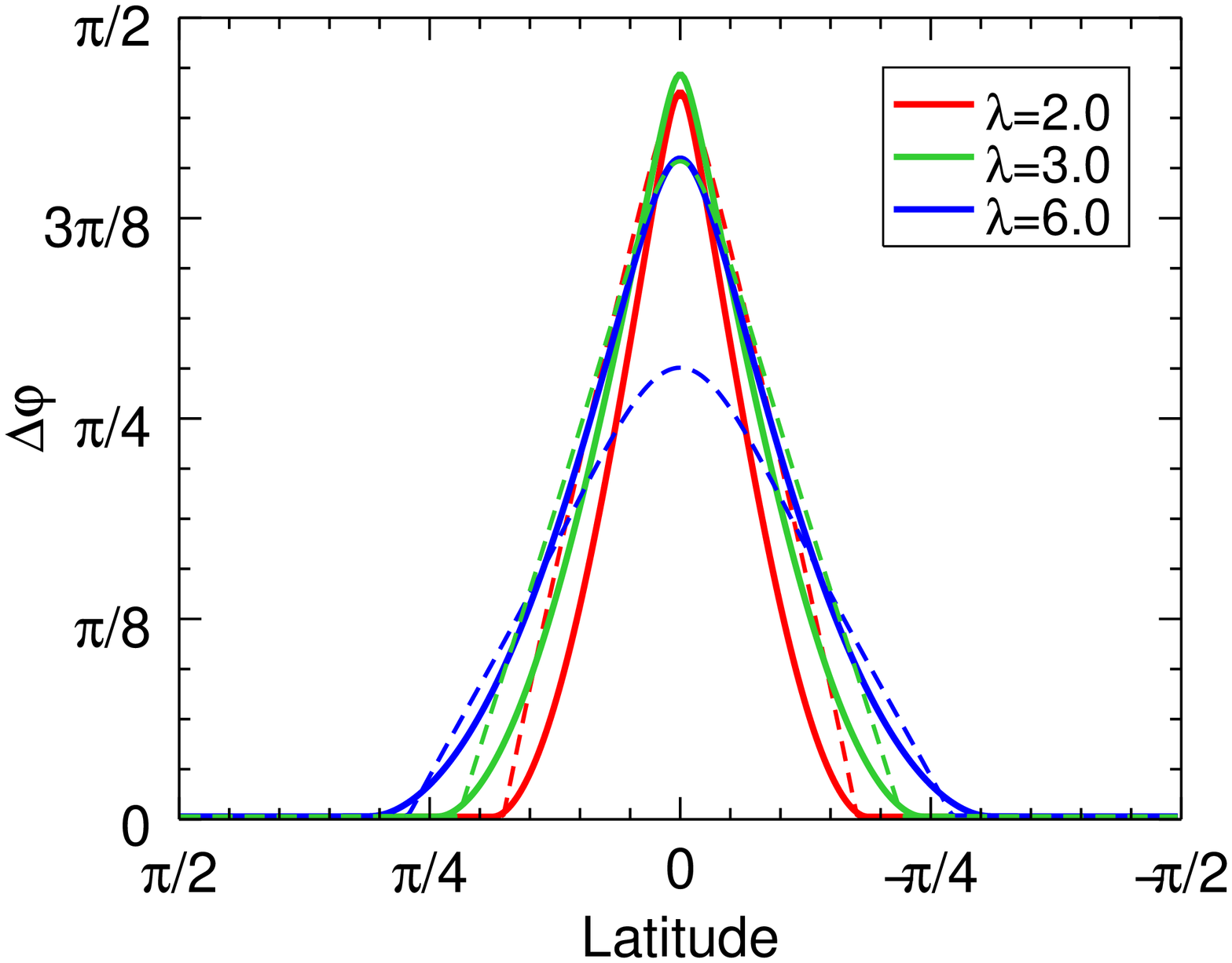}
	\caption{ Profiles of the twist angle $\Delta \varphi$ in the magnetosphere (left) and
			  at the stellar surface (right) for the same equilibrium configurations whose
			  magnetic field profiles have been shown in figure~\ref{fig:bprofZ1}. Solid
			  lines (dashed lines) refer to configurations with $\zeta=1$ ($\zeta=0$).
	          }
	\label{fig:dphiprofZ1}
\end{figure*}

The effects of the new current distribution on the twist angle $\Delta \varphi$ are shown 
in figure~\ref{fig:dphiprofZ1} where we plot the trend of $\Delta \phi$ 
for the same equilibria discussed above. 
While the growth of the surface $\Delta \phi$ is a direct consequence of the stronger 
toroidal field obtained for the $\zeta=1$ configurations, the analysis of trends in the magnetosphere
deserve more attention. In fact, even though the toroidal field in the new configurations
is stronger in the proximity of the stellar surface, the twist angle in the magnetosphere decreases
monotonically and it is highly suppressed with respect to that obtained in the $\zeta=0$ models.
This is due, on the one hand, to the fact that $B_{\rm tor}$ goes more rapidly to zero in the magnetosphere
but, on the other hand, also to the presence of a stronger equatorial poloidal field in the 
vicinity of the star.

\begin{figure*}
	\centering
	\includegraphics[width=.45\textwidth]{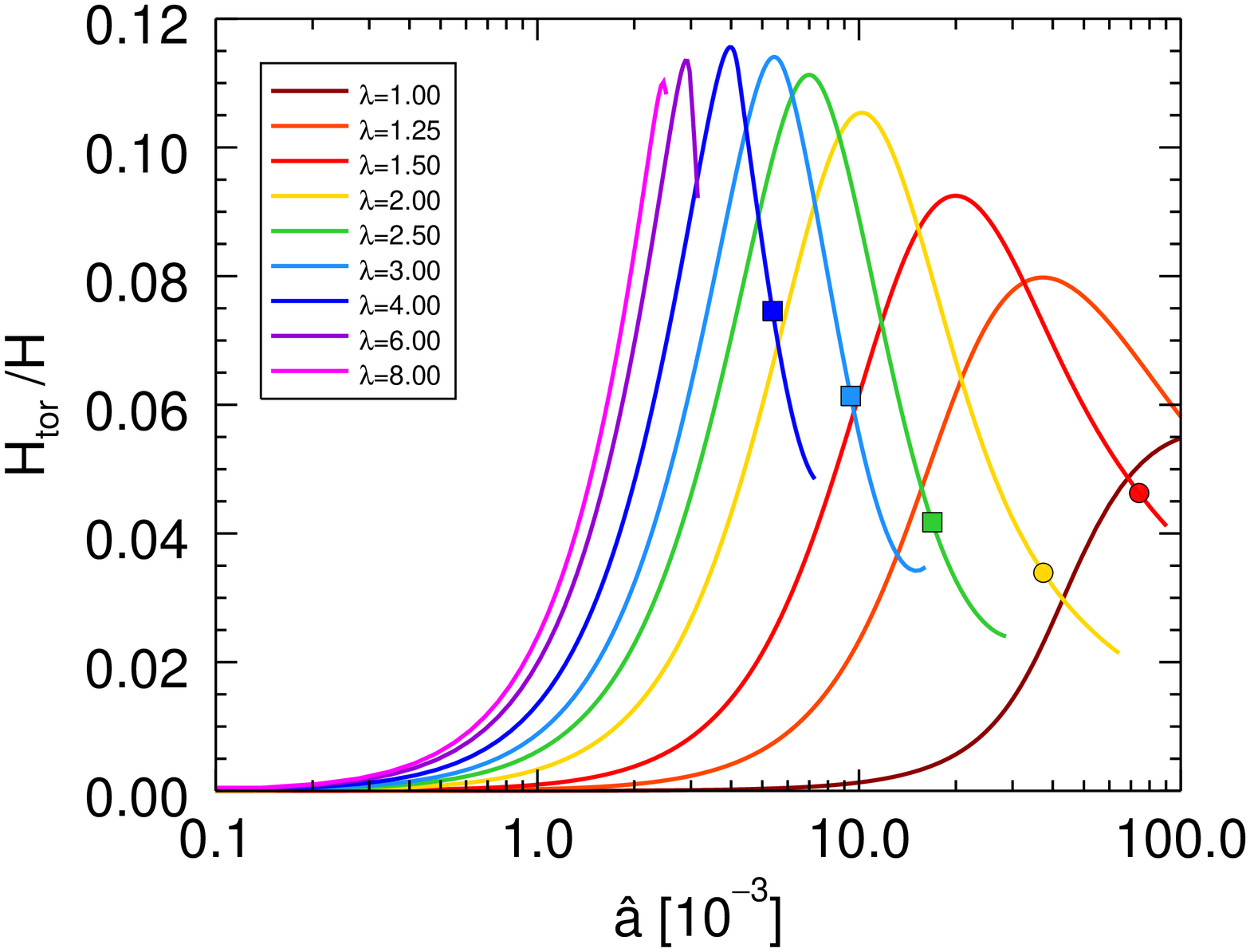} 
	\includegraphics[width=.45\textwidth]{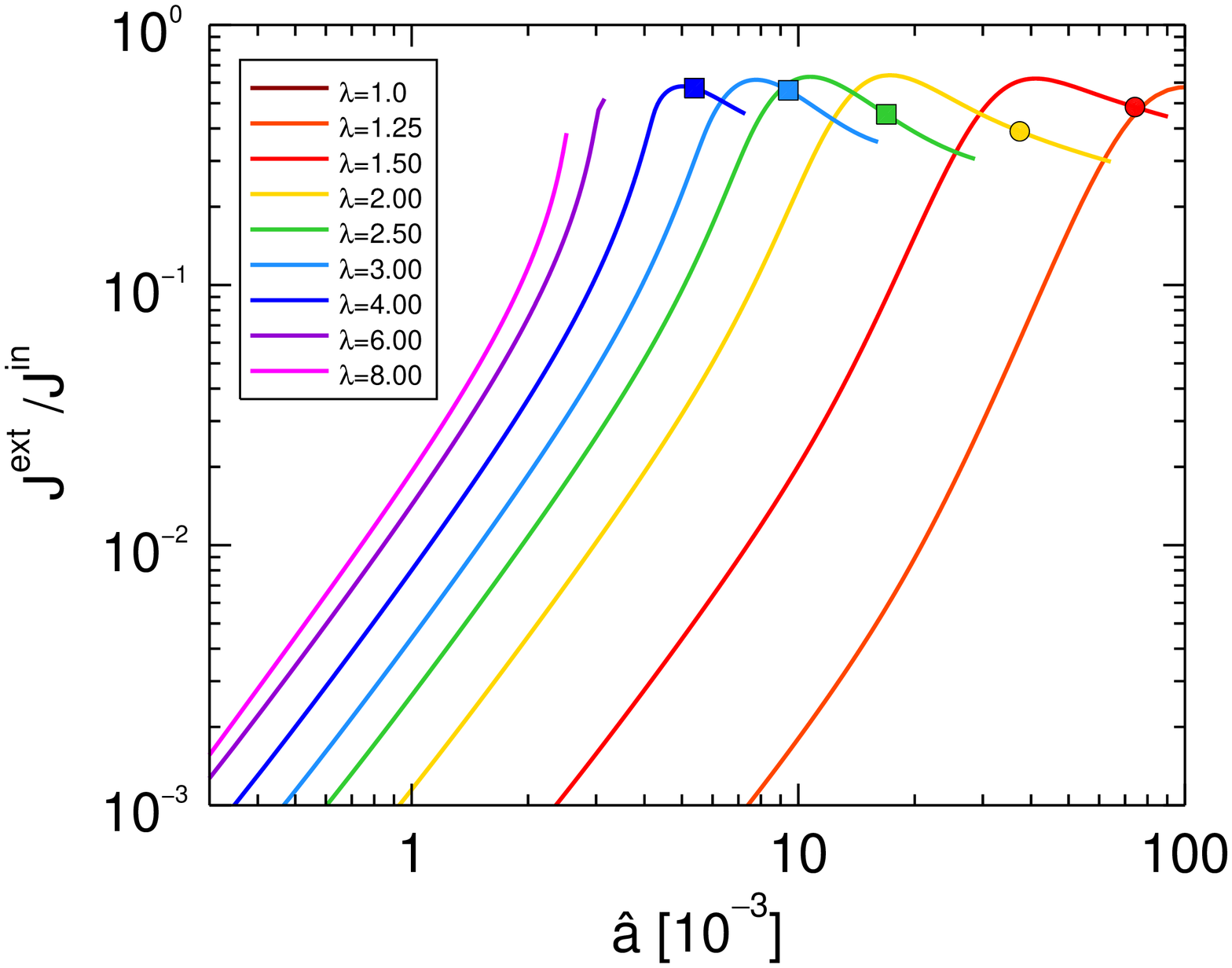} 
	\caption{ Profiles of the toroidal energy ratio  $\mathcal{H}_{\rm tor}/\mathcal{H}$ (left)
	and of the current ratio $\mathcal{J}^{\rm ext}/\mathcal{J}^{\rm in}$    
	as a function of $\hat{a}$ along equilibrium sequences with constant $\lambda$. 
	Dots and squares as in figure~\ref{fig:bmax}.
              }
	\label{fig:JoJZ1}
\end{figure*}

The structure of the magnetic field is however only slightly affected by the
value of the magnetic index $\zeta$. This is also evident from figure~\ref{fig:JoJZ1}
where we plot the profile of both the magnetic energy ratio
$\mathcal{H}_{ \rm tor}/\mathcal{H}$ and the current ratio $\mathcal{J}^{ \rm ext}/\mathcal{J}^{ \rm in}$
as a function of $\hat{a}$ for the various sequences. The trends strictly reflect those
obtained in the $\zeta=0$ case (see figure~\ref{fig:HoH} and~\ref{fig:joj}) and 
it is interesting to notice that sequences with equal $\lambda$ behave, from the point of
view of the field topology, in the same way:
for sequences with  $\lambda \lesssim 2$ the twisted region moves outside the star;
for sequences with  $\lambda \gtrsim 2.5$ the configuration at higher $\hat{a}$ are
characterized by a more complex topology and part of the toroidal field remain always
confined in the star. Also a more quantitative comparison shows little differences.
In the $\zeta=1$ case the maximum allowed  $\mathcal{H}_{ \rm tor}/\mathcal{H}$
is lower if $\lambda \lesssim 2.5$ and higher if $\lambda > 2.5$. The major differences 
regard the sequences with $\lambda=6$ and 8 where the higher value of $B_{\rm tor}$ and a
more regular topology of the solution (i.e. there is no formation of an X-point) allow to
reach higher value for the magnetic energy ratio. Finally in both cases, when 
$\mathcal{J}^{ \rm ext}\sim 0.7 \mathcal{J}^{ \rm in}$, the system self regulates
inducing a change in the topology of the distribution of the magnetic field and the 
associated external current.

\section{Conclusions}

\label{sec:conclusion}

There is an ever increasing amount of evidence that magnetars have a
strongly twisted magnetosphere, and that it is this twist more than
the strength of the field itself that define their phenomenology, and
isolate them as a separate class of NS. Investigating how this twisted
magnetosphere is arranged, and what could be its equilibrium
structure, is thus an important step for a more realistic description
of these astrophysical sources.

We have computed numerically, for the first time, equilibrium models of 
general relativistic magnetized NSs with twisted magnetospheres, 
allowing for electric currents extending smoothly from the interior
of the star to the exterior. Our work extends a recent study by
\citet{Glampedakis_Andersson+12a} in the Newtonian regime.

Our models represent a straightforward generalization of typical Twisted-Torus
configurations, where the twist is allowed to extend also outside the NS. In particular,
we have focused in the low-magnetization limit, since this limit is
appropriate for real physical system like AXPs and SGRs. In this case the morphology
of the magnetic field can be fully parametrized in terms of a single
quantity $\hat{a}$, independently of the strength of the magnetic
field. We have shown that the extent of the magnetosphere (our parameter
$\lambda$) plays an important role and defines the possible
existence of different topological classes of solutions.

In the low $\hat{a}$ regime, when the non-linear current terms are weak, 
the magnetic field lines are inflated outward by the toroidal magnetic field pressure and 
the twist of the field lines extends also to higher latitude. The
result is a single magnetically connected region.
As $\hat{a}$ increases, the effects of the non-linearity of the equation start to arise.
This not only reduces the twist of the near-surface magnetic field but also
leads to the formation of a disconnected magnetic island, reminiscent
of the so-called plasmoids often found in simulations of the solar corona. 
This regime and these topologies are very likely to be unstable.

In this work we focused on configurations with a magnetospheric confined twist exploring
the effects of different choices for the strength and shape of the twist.
Even though, as pointed out in \citet{Beloborodov11a}, the observations of 
shrinking hotspots on magnetar transient seem to suggest that the twist is more probably located 
near the pole, similar magnetospheric geometries have been recently used to model SGR giant flare
as flux rope eruption \citep{Huang_Yu14b,Huang_Yu14a}. 

Our approach to NS  magnetospheric equilibrium has allowed us to obtain complex magnetic field 
morphologies. However, apart from a rough estimate based on known
criteria, it is difficult to establish their stability, especially
with respect to non azimuthal perturbations. 
Moreover, since we treat the magnetosphere as force-free plasma, the
physical regime to which our
models apply, is characteristic of the late phases of a proto-NS, when a crust begins to form. Therefore, a meaningful
modellization of the evolution of the system can not disregard the important stabilizing role 
played by the crust. This is just indicative of the great complexity of
the physic involved, and correspondingly of the extreme difficulties in the realistic modelling of NS structure.

\section*{Acknowledgements}
This work has been supported by a EU FP7-CIG grant issued to the
NSMAG project (P.I. NB), and by the INFN TEONGRAV initiative 
(local P.I. LDZ).

\bibliography{my}{}
\bibliographystyle{mn2e}

\appendix
\section{Strong field regime}
\label{sec:app1}
Newly born magnetars, with their fast rotation (with period of the order of $\sim 1$ s) and
their strong magnetic deformation, can power a significant emission of gravitational waves and,
during the first few seconds of their life, they could be a promising target for the next
generation of ground-based GW-interferometers \citep{Mastrano_Melatos+11a}. 
After this lapse of time, because of the spin-down induced by the strong poloidal field ($\sim 10^{14}$ G)
the strain amplitude reduces considerably and the GW emission is
hardly detectable.

Even though our assumption of a force-free magnetosphere only applies to
the late phases of a proto-NS, when GW emission will be quenched,
it is still interesting to consider the strong field regime and how a
magnetospheric distribution of currents acts on the stellar deformation.
In the following we will limit our discussion to sequences with $\lambda=1.5$, considering
only configurations with a simple topology with no detached magnetic flux rope outside
the star. As discussed in section~\ref{sec:models} this kind of configurations,
whose properties are weakly affected by $\lambda$, are possibly the only stable ones.

\begin{figure}
	\centering
	\includegraphics[width=.45\textwidth]{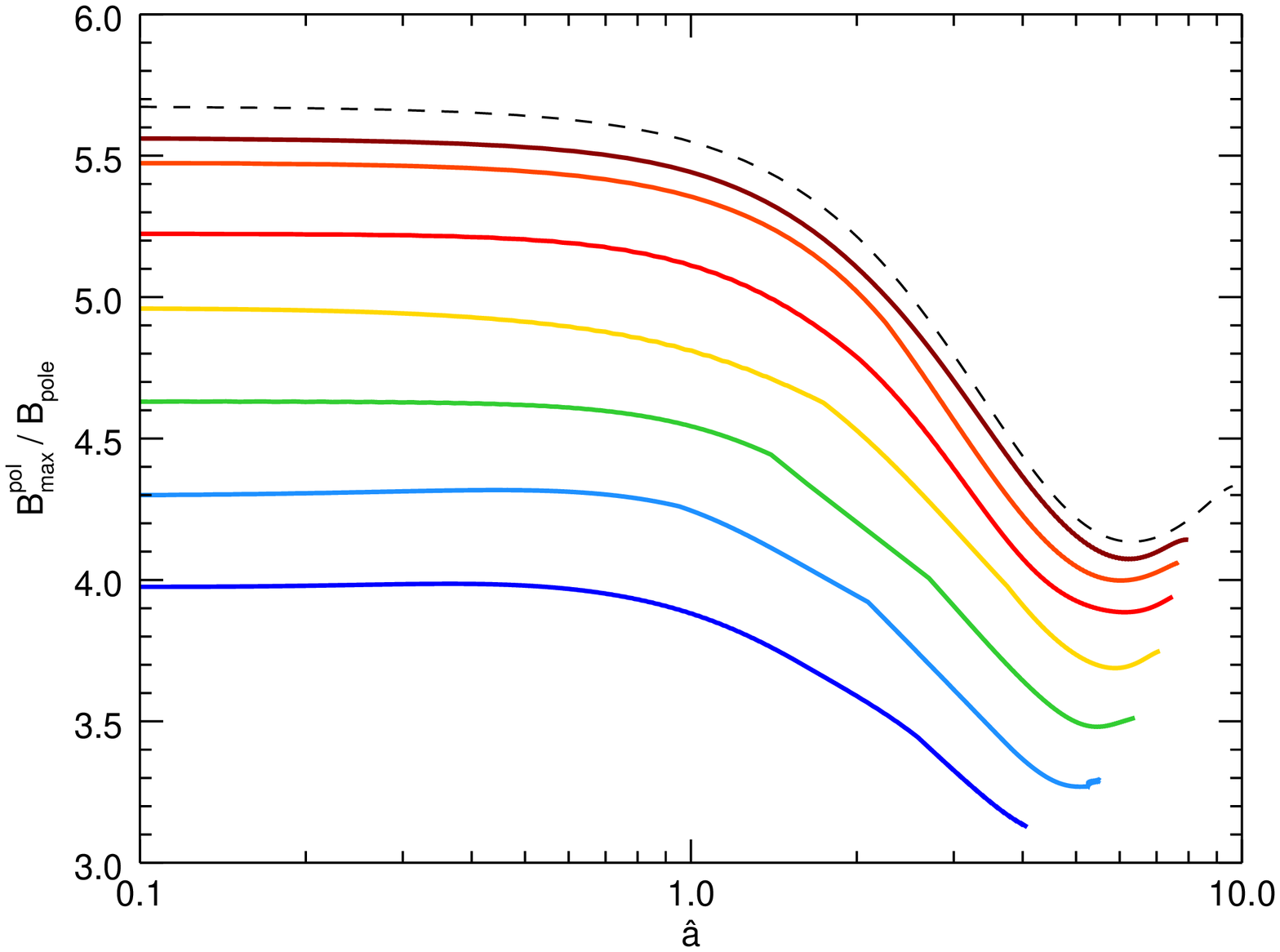} 
	\caption{ Profiles of the maximum value of the poloidal magnetic field $B_{\rm max}^{\rm pol}$
	along sequences with $\lambda=1.5$  and constant value of $B_{\rm pole}$. Solid lines from top
	to bottom $B_{\rm pole}=0.2,0.4,0.6,0.8,1.0,1.2,1.4 \times 10^{17}$ G. The black dashed line 
	corresponds to the weak field limit. 
              }
	\label{fig:Bstrong}
\end{figure}

In the strong field regime the solutions of the Grad-Shafranov equation~\ref{eq:gs} depend
on the specific value of the magnetic field strength and they do not rescale as in 
the weak-field limit. This is evident from figure~\ref{fig:Bstrong} where  we show the value of the ratio 
$B_{\rm max}^{\rm pol}/B_{\rm pole}$ as a function of $\hat{a}$ along sequences with 
constant value of $B_{\rm pole}$ (from the weak field limit to $\sim 10^{17}$ G) and fixed 
gravitational mass $M=1.551 M_{\sun}$. Just like in the weak-field regime, along each sequence the trend of 
$B_{\rm max}^{\rm pol}$  is not monotonic. Here, however, the ratio $B_{\rm max}^{\rm pol}/B_{\rm pole}$ 
is smaller for stronger magnetic fields. This can be explained in term of the deformation of the 
star: as in the weak-field limit our equilibrium configurations are energetically dominated by the poloidal 
field (the magnetic energy ratio $\mathcal{H}_{\rm tor}/\mathcal{H}$
depends weakly on the strength of the field) and they show an oblate
deformation. Therefore, if the star is more magnetized, the deformation
is stronger and the pole is closer to the center of the star implying a smaller 
$B_{\rm max}^{\rm pol}/B_{\rm pole}$.

\begin{figure}
	\centering
	\includegraphics[width=.45\textwidth]{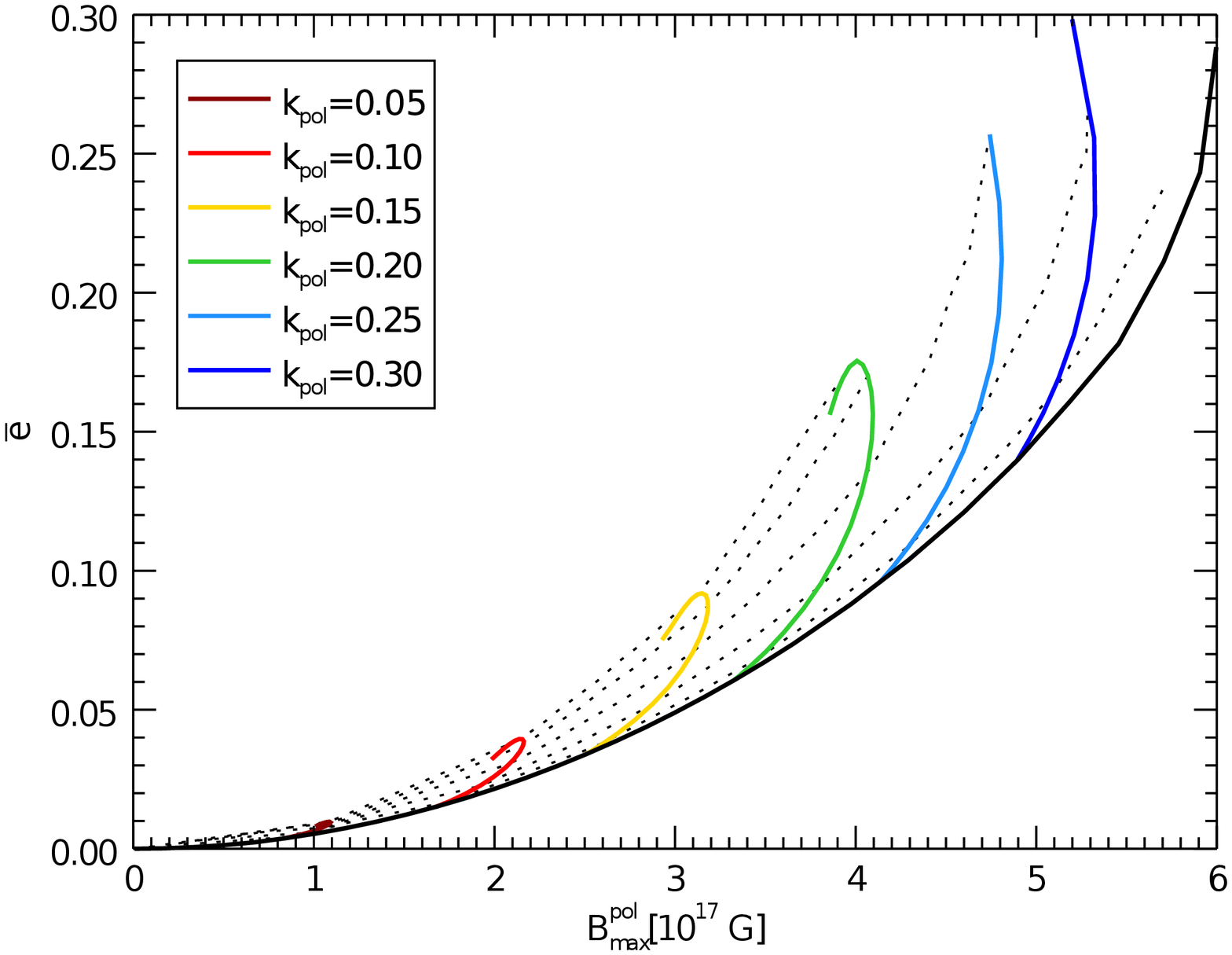} 
	\caption{ Profiles of the deformation rate $\bar{e}$ along sequences with constant value of 
	$k_{\rm pol}$, $\lambda=1.5$, and gravitational mass $M=1.551 M_{\sun}$ as a function of 
	the maximum value of the poloidal magnetic field $B_{\rm max}^{\rm pol}$. The black solid line 
	represents the sequence with $\hat{a}=0$ (purely poloidal configurations). The dotted lines 
	correspond to sequences with	fixed value of $\hat{a}$ equispaced from 
	$\hat{a}=1.0\times 10^{-3}$ to $\hat{a}=5.0\times 10^{-3}$.
              }
	\label{fig:def}
\end{figure}

Figure~\ref{fig:def} shows the deformation rate $\bar{e}$ \citep{Kiuchi_Yoshida08a,Pili_Bucciantini+14a} as 
a function of $B_{\rm max}^{\rm pol}$.
Here the black lines trace the configurations with constant $\hat{a}$.
Moving from the purely poloidal case with $\hat{a}=0$  to higher value 
of $\hat{a}$  the toroidal magnetic field strengthen up to $\sim 2 \times 10^{17}$ G
and the deformation rate increases by a factor $\sim 2$. The presence of a toroidal field to the system, 
seems to increase, rather then reduce, the oblate deformation. However, as discussed in PBD14,
neither the maximum strength of the magnetic field nor the magnetic energy are, in general, good 
indicator of the possible deformations of the NS. The distribution of the currents play 
an important role and the effects of magnetic field located in the outer layers of the star are less 
important than those of comparable field situated in the core region.
Therefore the trend of $\bar{e}$ can be explained in terms of the strength of the poloidal magnetic field,
which resides deeper inside the star. Moving along sequences with fixed $k_{\rm pol}$ (without constraints 
on $B_{\rm pole}$) from the purely poloidal configurations to higher $\hat{a}$, both the poloidal and the 
toroidal field grow in strength while the neutral line moves toward the surface of the star. Finally,
as soon as the radius of the neutral line reaches a value of $\sim 0.8 r_{\rm e}$, the poloidal field begin 
to drop leading to an inversion point in the sequences and a reduction of the deformation rate $\bar{e}$.

The deformation rate $\bar{e}$ has the advantage that it can be computed as an integral over the star, but 
it is strictly a Newtonian quantity. In general relativity the the relevant quantity for the emission of 
gravitational waves is the quadrupolar ellipticity $e_{\rm q}$ defined as:
\be
e_{\rm q}=-\frac{3}{2}\frac{I_{zz}}{I},
\ee  
where $I_{zz}$ is the gravitational quadrupole moment and $I$ is the moment of inertia.  
The gravitational quadrupole moment  can not be computed as an integral over the star
but must be derived from the asymptotic structure of the metric terms \citep{Bonazzola_Gourgoulhon96b}
in the limit $r\rightarrow\infty$. The moment of inertia can be properly defined only for rotating star  as 
the ratio of the Komar angular momentum \citep{Kiuchi_Yoshida08a} over the rotational rate 
$I:=\mathscr{J}/\Omega$  \citep{Bonazzola_Gourgoulhon96b,Frieben_Rezzolla12a}. For non rotating stars it can 
be evaluated in the limit $\Omega\rightarrow0$:
\be
I= \lim_{\Omega\rightarrow 0} \frac{\mathscr{J}}{\Omega}=\int(e+p) \psi^{10}\alpha^{-1}r^4 \sin^3\theta dr 
d\theta d\phi.
\ee
We find that in all our models the quadrupolar ellipticity is an almost constant fraction of
the deformation rate: $e_{\rm q}/\bar{e}= 0.40 \pm 0.05$ (the uncertainty is mostly due to the asymptotic
extrapolation of the metric terms). This agrees with what was already found by \citet{Frieben_Rezzolla12a}.
\end{document}